\documentclass[aps,prd,preprint,tightenlines,superscriptaddress,showpacs,byrevtex,preprintnumbers,amsmath,amssymb]{revtex4-1}

\usepackage{graphicx} % Include figure files
\usepackage{dcolumn}  % Align table columns on decimal point
\usepackage{color}  % Align table columns on decimal point

\graphicspath{{ps}}

\begin{document}

\vspace*{-3\baselineskip}
\resizebox{!}{3cm}{\includegraphics{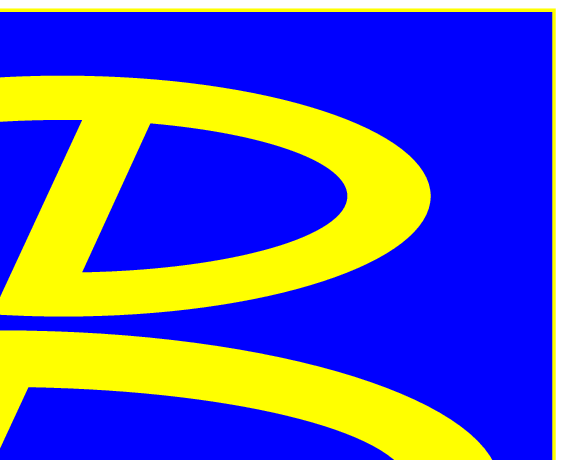}}

\preprint{\vbox{
\hbox{Belle Preprint 2014-16 }
\hbox{KEK Preprint 2014-27}
}}
%\vspace*{3\baselineskip}

\title{ \quad\\[1.0cm]  Measurement of \boldmath $B^0 \to D_s^- K^0_S\pi^+$ and $B^+ \to D_s^- K^+K^+$ branching fractions}

\begin{abstract}
We report a measurement of the $B^0$ and $B^+$ meson decays to the $D_s^-K^0_S \pi^+$  and $D_s^- K^+K^+$ final states, respectively, using $657 \times 10^{6} B\overline{B}$ pairs
collected at the $\Upsilon(4S)$ resonance with the Belle detector
at the KEKB asymmetric-energy $e^+e^-$ collider.
Using the $D_s^-\to \phi\pi^-$, $K^{*}(892)^0 K^-$ and
$K^0_S K^-$ decay modes for the $D_s$ reconstruction, we measure the 
following branching fractions:
${\cal B}(B^0\to D_s^{-}K^0_S\pi^+)=
[0.47 \pm 0.06 (\mathrm {stat}) \pm 0.05 (\mathrm {syst})]\times 10^{-4}$ and ${\cal B}(B^+\to D_s^-K^+K^+)=
[0.93 \pm 0.22 (\mathrm {stat})\pm 0.10 (\mathrm {syst})]\times 10^{-5}$. We find the ratio of the branching fraction of $B^+\to D_s^-K^+K^+$ to that of the analogous Cabibbo favored $B^+\to D_s^-K^+\pi^+$ decay to be ${\cal R}_{\cal B} = 0.054 \pm 0.013 ({\rm stat}) \pm 0.006 ({\rm syst})$, which is consistent with the na\"{\i}ve factorization model. We also observe a deviation from the three-body phase-space model for both studied decays.
\end{abstract}


%%%% >>>> keep the final version single-spaced
%\tighten

\affiliation{University of the Basque Country UPV/EHU, 48080 Bilbao}
\affiliation{Beihang University, Beijing 100191}
\affiliation{University of Bonn, 53115 Bonn}
\affiliation{Budker Institute of Nuclear Physics SB RAS and Novosibirsk State University, Novosibirsk 630090}
\affiliation{Faculty of Mathematics and Physics, Charles University, 121 16 Prague}
%%%\affiliation{Chiba University, Chiba 263-8522}
\affiliation{Chonnam National University, Kwangju 660-701}
\affiliation{University of Cincinnati, Cincinnati, Ohio 45221}
\affiliation{Deutsches Elektronen--Synchrotron, 22607 Hamburg}
%%%\affiliation{Department of Physics, Fu Jen Catholic University, Taipei 24205}
\affiliation{Justus-Liebig-Universit\"at Gie\ss{}en, 35392 Gie\ss{}en}
\affiliation{Gifu University, Gifu 501-1193}
%%%\affiliation{II. Physikalisches Institut, Georg-August-Universit\"at G\"ottingen, 37073 G\"ottingen}
\affiliation{The Graduate University for Advanced Studies, Hayama 240-0193}
\affiliation{Gyeongsang National University, Chinju 660-701}
\affiliation{Hanyang University, Seoul 133-791}
\affiliation{University of Hawaii, Honolulu, Hawaii 96822}
\affiliation{High Energy Accelerator Research Organization (KEK), Tsukuba 305-0801}
%%%\affiliation{Hiroshima Institute of Technology, Hiroshima 731-5193}
\affiliation{IKERBASQUE, Basque Foundation for Science, 48011 Bilbao}
%%%\affiliation{University of Illinois at Urbana-Champaign, Urbana, Illinois 61801}
%%%\affiliation{Indian Institute of Technology Bhubaneswar, Satya Nagar 751007}
\affiliation{Indian Institute of Technology Guwahati, Assam 781039}
\affiliation{Indian Institute of Technology Madras, Chennai 600036}
\affiliation{Indiana University, Bloomington, Indiana 47408}
\affiliation{Institute of High Energy Physics, Chinese Academy of Sciences, Beijing 100049}
\affiliation{Institute of High Energy Physics, Vienna 1050}
%%%\affiliation{Institute for High Energy Physics, Protvino 142281}
%%%\affiliation{Institute of Mathematical Sciences, Chennai 600113}
\affiliation{INFN - Sezione di Torino, 10125 Torino}
\affiliation{Institute for Theoretical and Experimental Physics, Moscow 117218}
\affiliation{J. Stefan Institute, 1000 Ljubljana}
\affiliation{Kanagawa University, Yokohama 221-8686}
\affiliation{Institut f\"ur Experimentelle Kernphysik, Karlsruher Institut f\"ur Technologie, 76131 Karlsruhe}
%%%\affiliation{Kavli Institute for the Physics and Mathematics of the Universe (WPI), University of Tokyo, Kashiwa 277-8583}
\affiliation{Kennesaw State University, Kennesaw GA 30144}
\affiliation{Department of Physics, Faculty of Science, King Abdulaziz University, Jeddah 21589}
\affiliation{Korea Institute of Science and Technology Information, Daejeon 305-806}
\affiliation{Korea University, Seoul 136-713}
%%%\affiliation{Kyoto University, Kyoto 606-8502}
\affiliation{Kyungpook National University, Daegu 702-701}
\affiliation{\'Ecole Polytechnique F\'ed\'erale de Lausanne (EPFL), Lausanne 1015}
\affiliation{Faculty of Mathematics and Physics, University of Ljubljana, 1000 Ljubljana}
\affiliation{Luther College, Decorah, Iowa 52101}
\affiliation{University of Maribor, 2000 Maribor}
\affiliation{Max-Planck-Institut f\"ur Physik, 80805 M\"unchen}
\affiliation{School of Physics, University of Melbourne, Victoria 3010}
\affiliation{Moscow Physical Engineering Institute, Moscow 115409}
\affiliation{Moscow Institute of Physics and Technology, Moscow Region 141700}
\affiliation{Graduate School of Science, Nagoya University, Nagoya 464-8602}
\affiliation{Kobayashi-Maskawa Institute, Nagoya University, Nagoya 464-8602}
%%%\affiliation{Nara University of Education, Nara 630-8528}
\affiliation{Nara Women's University, Nara 630-8506}
%%%\affiliation{National Central University, Chung-li 32054}
\affiliation{National United University, Miao Li 36003}
\affiliation{Department of Physics, National Taiwan University, Taipei 10617}
\affiliation{H. Niewodniczanski Institute of Nuclear Physics, Krakow 31-342}
%%%\affiliation{Nippon Dental University, Niigata 951-8580}
\affiliation{Niigata University, Niigata 950-2181}
%%%\affiliation{University of Nova Gorica, 5000 Nova Gorica}
\affiliation{Osaka City University, Osaka 558-8585}
%%%\affiliation{Osaka University, Osaka 565-0871}
\affiliation{Pacific Northwest National Laboratory, Richland, Washington 99352}
%%%\affiliation{Panjab University, Chandigarh 160014}
\affiliation{Peking University, Beijing 100871}
\affiliation{University of Pittsburgh, Pittsburgh, Pennsylvania 15260}
%%%\affiliation{Punjab Agricultural University, Ludhiana 141004}
%%%\affiliation{Research Center for Electron Photon Science, Tohoku University, Sendai 980-8578}
%%%\affiliation{Research Center for Nuclear Physics, Osaka University, Osaka 567-0047}
%%%\affiliation{RIKEN BNL Research Center, Upton, New York 11973}
%%%\affiliation{Saga University, Saga 840-8502}
\affiliation{University of Science and Technology of China, Hefei 230026}
\affiliation{Seoul National University, Seoul 151-742}
%%%\affiliation{Shinshu University, Nagano 390-8621}
\affiliation{Soongsil University, Seoul 156-743}
\affiliation{Sungkyunkwan University, Suwon 440-746}
\affiliation{School of Physics, University of Sydney, NSW 2006}
\affiliation{Department of Physics, Faculty of Science, University of Tabuk, Tabuk 71451}
\affiliation{Tata Institute of Fundamental Research, Mumbai 400005}
\affiliation{Excellence Cluster Universe, Technische Universit\"at M\"unchen, 85748 Garching}
\affiliation{Toho University, Funabashi 274-8510}
%%%\affiliation{Tohoku Gakuin University, Tagajo 985-8537}
\affiliation{Tohoku University, Sendai 980-8578}
\affiliation{Department of Physics, University of Tokyo, Tokyo 113-0033}
\affiliation{Tokyo Institute of Technology, Tokyo 152-8550}
\affiliation{Tokyo Metropolitan University, Tokyo 192-0397}
%%%\affiliation{Tokyo University of Agriculture and Technology, Tokyo 184-8588}
\affiliation{University of Torino, 10124 Torino}
%%%\affiliation{Toyama National College of Maritime Technology, Toyama 933-0293}
\affiliation{CNP, Virginia Polytechnic Institute and State University, Blacksburg, Virginia 24061}
\affiliation{Wayne State University, Detroit, Michigan 48202}
\affiliation{Yamagata University, Yamagata 990-8560}
\affiliation{Yonsei University, Seoul 120-749}
  \author{J.~Wiechczynski}\affiliation{H. Niewodniczanski Institute of Nuclear Physics, Krakow 31-342} % Krakow
  \author{J.~Stypula}\affiliation{H. Niewodniczanski Institute of Nuclear Physics, Krakow 31-342} % Krakow
  \author{A.~Abdesselam}\affiliation{Department of Physics, Faculty of Science, University of Tabuk, Tabuk 71451} % Tabuk
  \author{I.~Adachi}\affiliation{High Energy Accelerator Research Organization (KEK), Tsukuba 305-0801}\affiliation{The Graduate University for Advanced Studies, Hayama 240-0193} % KEK
  \author{K.~Adamczyk}\affiliation{H. Niewodniczanski Institute of Nuclear Physics, Krakow 31-342} % Krakow
  \author{H.~Aihara}\affiliation{Department of Physics, University of Tokyo, Tokyo 113-0033} % Tokyo
  \author{S.~Al~Said}\affiliation{Department of Physics, Faculty of Science, University of Tabuk, Tabuk 71451}\affiliation{Department of Physics, Faculty of Science, King Abdulaziz University, Jeddah 21589} % Tabuk
  \author{K.~Arinstein}\affiliation{Budker Institute of Nuclear Physics SB RAS and Novosibirsk State University, Novosibirsk 630090} % BINP
% \author{Y.~Arita}\affiliation{Graduate School of Science, Nagoya University, Nagoya 464-8602} % Nagoya
  \author{D.~M.~Asner}\affiliation{Pacific Northwest National Laboratory, Richland, Washington 99352} % PNNL
% \author{T.~Aso}\affiliation{Toyama National College of Maritime Technology, Toyama 933-0293} % Toyama
  \author{V.~Aulchenko}\affiliation{Budker Institute of Nuclear Physics SB RAS and Novosibirsk State University, Novosibirsk 630090} % BINP
% ...
% Ljubljana
% \author{N.~Zwahlen}\affiliation{\'Ecole Polytechnique F\'ed\'erale de Lausanne (EPFL), Lausanne 1015} % Lausanne
% \author{O.~Zyukova}\affiliation{Budker Institute of Nuclear Physics SB RAS and Novosibirsk State University, Novosibirsk 630090} % BINP
\collaboration{The Belle Collaboration}

\maketitle

{\renewcommand{\thefootnote}{\fnsymbol{footnote}}}
\setcounter{footnote}{0}

\begin{figure*}[tbh]
\begin{minipage}[b]{.46\linewidth}
\centering
\setlength{\unitlength}{1mm}
\includegraphics[height=3.7cm,width=7.4cm]{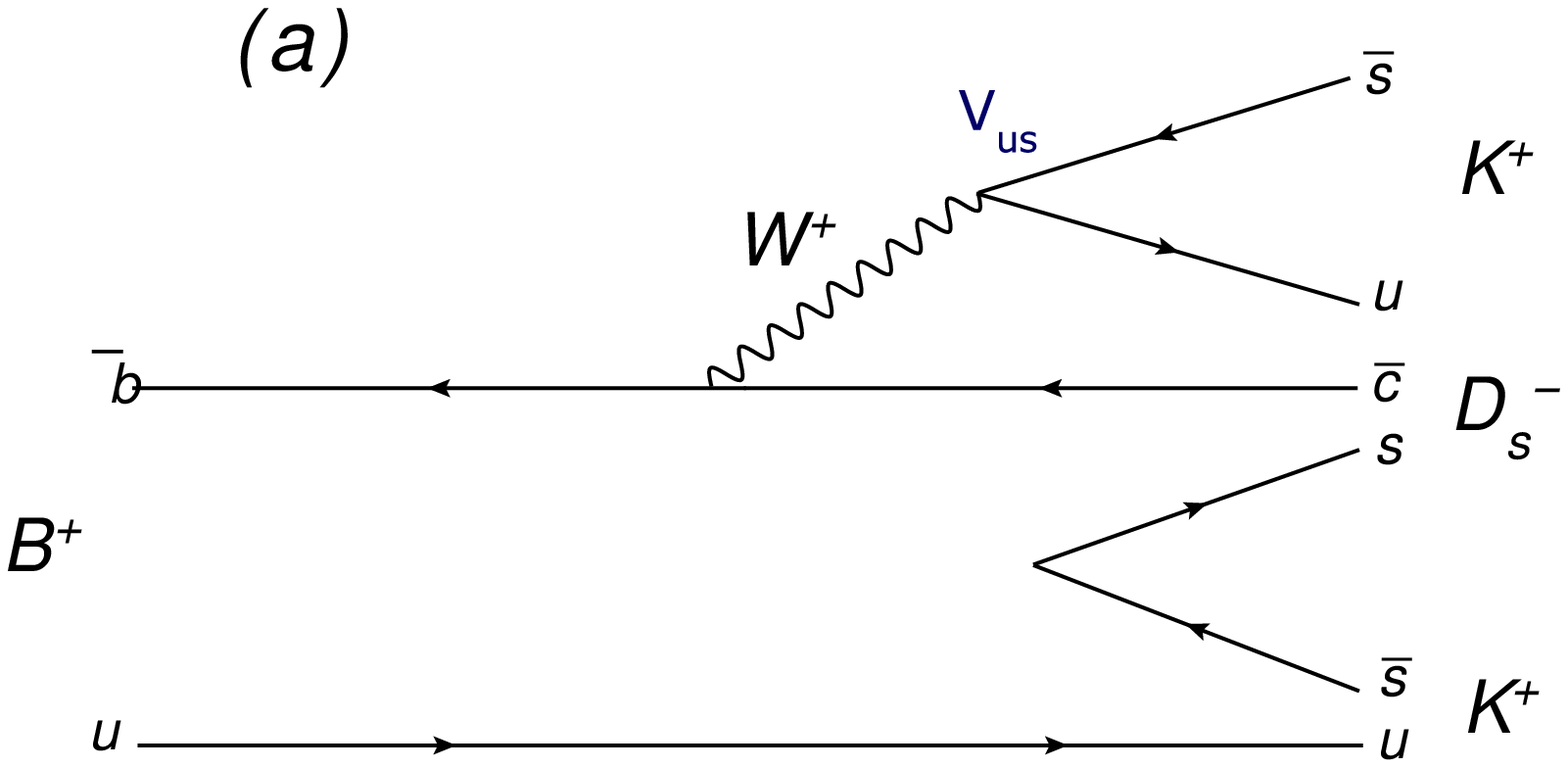}
\end{minipage}\hfill
\begin{minipage}[b]{.46\linewidth}
\centering
\setlength{\unitlength}{1mm}
\includegraphics[height=3.7cm,width=7.4cm]{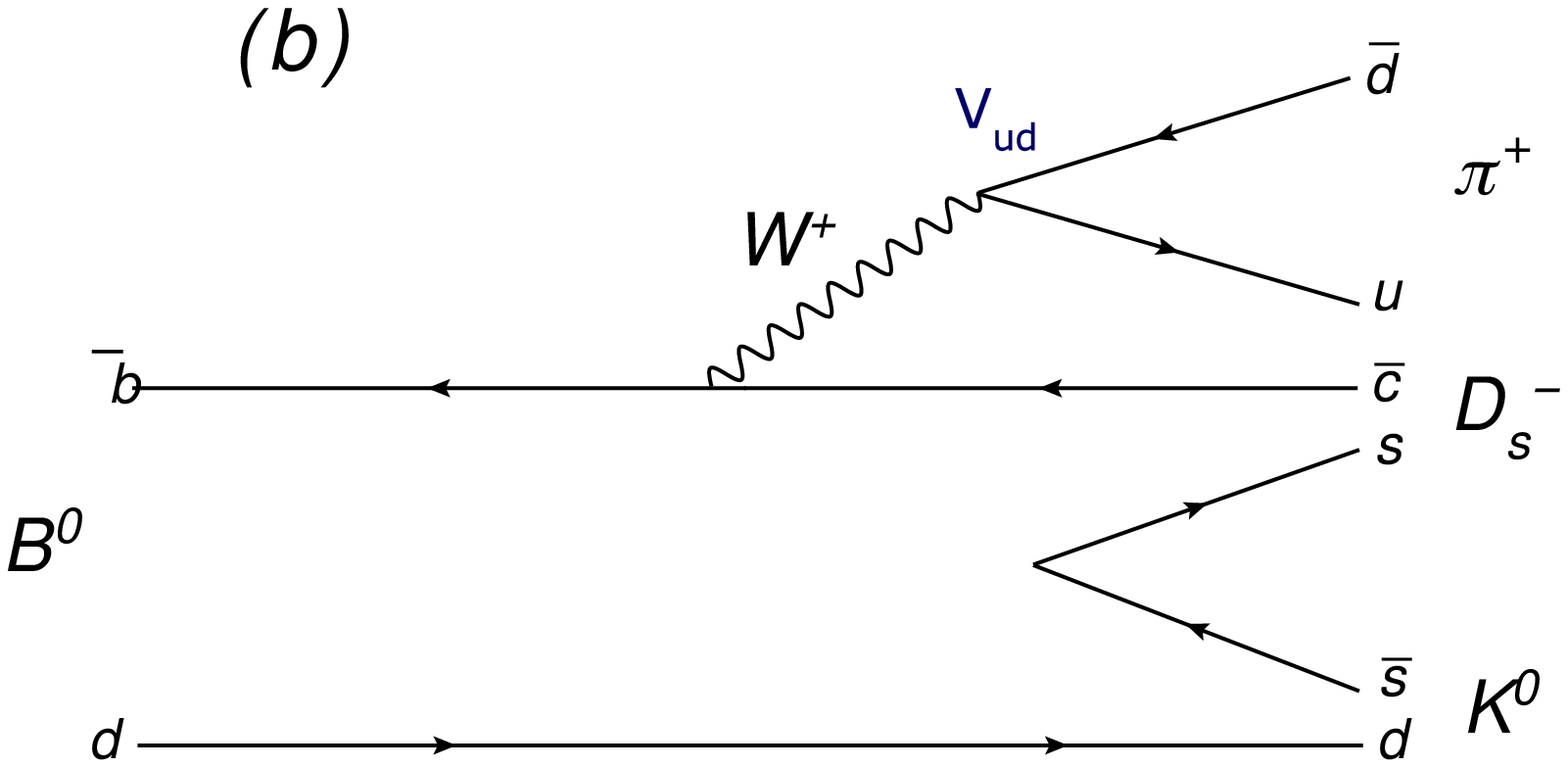}
\end{minipage}
\caption{ Dominant Feynman diagram for the (a) $B^+ \to D_s^- K^+K^+$ and (b) $B^0\to D_s^- K^0_S\pi^+$ decay.
}
\label{diagrams}
\end{figure*}

The dominant process for the decays  $B^0 \to D_s^-K^0_S \pi^+$ and $B^+ \to D_s^-K^+K^+$~\cite{FOOT} is mediated by the
$b\to c$ quark transition with subsequent $W$ fragmentation to a charged pion or kaon and includes the production of an additional $s\overline{s}$ pair, as shown in Fig.~\ref{diagrams}.
As the process $B^+ \to D_s^- K^+K^+$ is Cabibbo suppressed due to the formation of a $u\bar{s}$ pair from the $W$ vertex (Fig.~\ref{diagrams}a), its branching fraction can be compared
to the measured branching fraction of the Cabibbo favored
$B^+ \to D_s^-K^+\pi^+$ decay~\cite{MOJE, BABAR_DSKAPI}.
Within the framework of na\"{\i}ve factorization~\cite{FACTOR},
the ratio of these branching fractions should be proportional to
the ratio of the squares of the CKM matrix elements $V_{ud}$ and $V_{us}$~\cite{CKM_PAPER1, CKM_PAPER2}. 
Such a comparison allows us to check the validity of existing theoretical descriptions of the three-body hadronic decays.
In addition, the two-body subsystem of the $D_s^- K^0_S\pi^+$ and $D_s^- K^+K^+$ final states merits study since a significant deviation from the simple phase-space model was observed in the $D_s^-K^+$ invariant mass for the similar process $B^+ \to D_s^- K^+\pi^+$~\cite{MOJE, BABAR_DSKAPI} and also in the semileptonic process $B^+ \to D_s^{(*)-}K^+l^+\nu_l$~\cite{JACO}. This constitutes a potential source of new spectroscopy discoveries.

Both $B^0 \to D_s^-K^0_S \pi^+$ and $B^+ \to D_s^-K^+K^+$ decay modes have been observed by BaBar~\cite{BABAR_DSKAPI} and call for confirmation.
In this paper, we report measurements of the branching fractions  for $B^0 \to D_s^-K^0_S \pi^+$ and
$B^+ \to D_s^-K^+K^+$ and compare the latter's with the branching fraction for $B^+ \to D_s^- K^+\pi^+$. The invariant mass distributions for the two-body subsystems are studied to evaluate the discrepancy from the phase-space model. The analysis is performed on a data sample containing $(657\pm 9)\times 10^6$ $B\overline{B}$ pairs collected  
with the  Belle detector at the KEKB asymmetric-energy $e^+e^-$ collider~\cite{KEKB}
operating at the $\Upsilon(4S)$ resonance.
The production rates of $B^+B^-$ and $B^0\overline{B}{}^0$ pairs are assumed to be equal.

The Belle detector~\cite{BELLE} is a large-solid-angle magnetic spectrometer that
consists of a silicon vertex detector (SVD), a 50-layer central drift
chamber (CDC), an array of aerogel threshold Cherenkov counters
(ACC), a barrel-like arrangement of time-of-flight scintillation
counters (TOF), and an electromagnetic calorimeter composed of CsI(Tl)
crystals, all located inside a superconducting solenoid coil that
provides a 1.5~T magnetic field. An iron flux return located outside
of the coil is instrumented to detect $K_L^0$ mesons and to identify
muons. Two inner detector configurations were
used: a 2.0~cm beam pipe with a 3-layer SVD for the first sample of $152 \times 10^6 B\overline{B}$ pairs and 
a 1.5~cm beam pipe with a 4-layer SVD for the remaining $505 \times 10^6 B\overline{B}$
pairs \cite{SVD}.

Charged tracks are required to have a distance of closest approach 
to the interaction point of less than 5.0~cm along the positron beam direction (defined to be the $z$-axis)
and less than 0.5~cm in the transverse plane.
In addition, charged tracks must have transverse momenta larger than 
 $100~{\rm MeV}/c$. To identify charged hadrons, we combine information from 
 the CDC, ACC and TOF into pion, kaon and proton likelihoods
${\cal L}_{\pi}$, ${\cal L}_{K}$ and ${\cal L}_{p}$, respectively.
 For a kaon candidate, we require the likelihood ratio
${\cal L}_{K/\pi} = {\cal L}_K/({\cal L}_K + {\cal L}_\pi)$ to be greater than 0.6.
Pions are selected from track candidates with low kaon probabilities satisfying
${\cal L}_{K/\pi} < 0.95$. For kaons (pions), we also apply a proton veto criterion: ${\cal L}_{p/K} ({\cal L}_{p/\pi}) < 0.95$. In addition, we reject all charged tracks consistent with an electron (or muon)
hypothesis ${\cal L}_{e(\mu)} < 0.95$, where ${\cal L}_{e}$ and ${\cal L}_{\mu}$ are respective lepton likelihoods. The above requirements result in a typical momentum-dependent kaon (pion) identification efficiency
ranging from 92\% to 97\% (94\% to 98\%) for various channels,
with 2-15\% of kaon candidates being misidentified as pions and
4-8\% of pion candidates being misidentified as kaons.

The $D_s^-$ candidates are reconstructed in three final states:
$\phi(\to K^+K^-)\pi^- $,  $K^*(892)^0(\to K^+\pi^-) K^- $ and $K^0_S(\to \pi^+ \pi^-) K^- $.
We retain $K^+K^-$  ($K^+\pi^-$) pairs  as $\phi$ ($K^{*}(892)^0$) candidates if their invariant mass lies  within 10 (100) MeV/$c^2$ of the nominal $\phi~(K^{*}(892)^0 )$ mass~\cite{PDG}.
This requirement has 91\% (95\%) efficiency for the respective $D_s$ decay mode. Candidate $K^0_S$ mesons are selected by combining pairs of oppositely charged tracks (treated as pions) with an invariant
mass within 16~MeV/$c^2$ ($3\sigma$) of the nominal $K^0_S$ mass. In addition, the vertices of these 
track pairs must be displaced from the interaction point by at least 0.5~cm.

\begin{figure*}[tbh]
\begin{minipage}[b]{.32\linewidth}
\centering
\setlength{\unitlength}{1mm}
\begin{picture}(60,50)
\includegraphics[height=4.9cm,width=5.9cm]{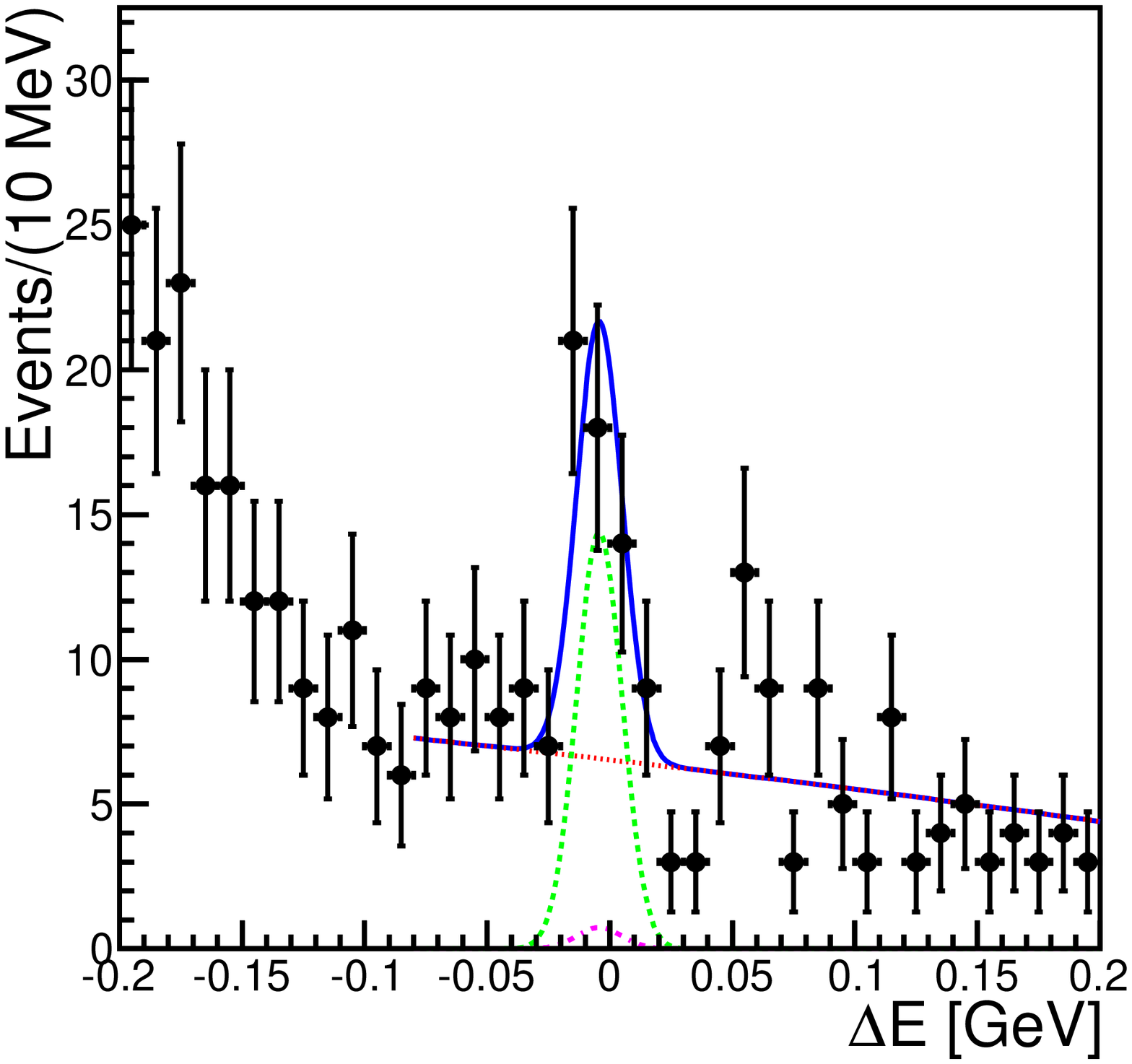}
\end{picture}
\end{minipage}\hfill
\begin{minipage}[b]{.32\linewidth}
\centering
\setlength{\unitlength}{1mm}
\begin{picture}(60,50)
\includegraphics[height=4.9cm,width=5.9cm]{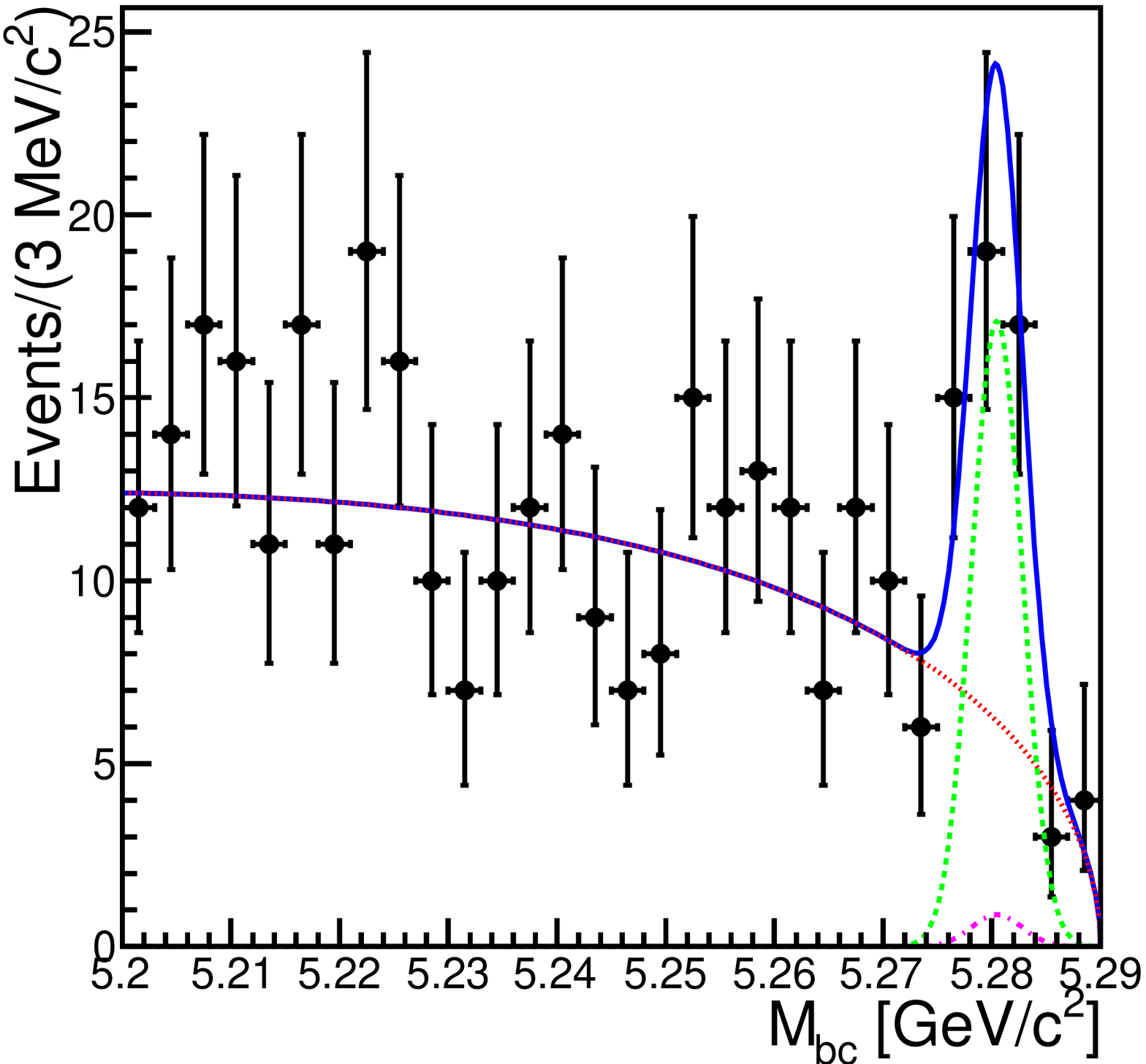}
\end{picture}
\end{minipage}\hfill
\begin{minipage}[b]{.32\linewidth}
\centering
\setlength{\unitlength}{1mm}
\begin{picture}(60,50)
\includegraphics[height=4.9cm,width=5.9cm]{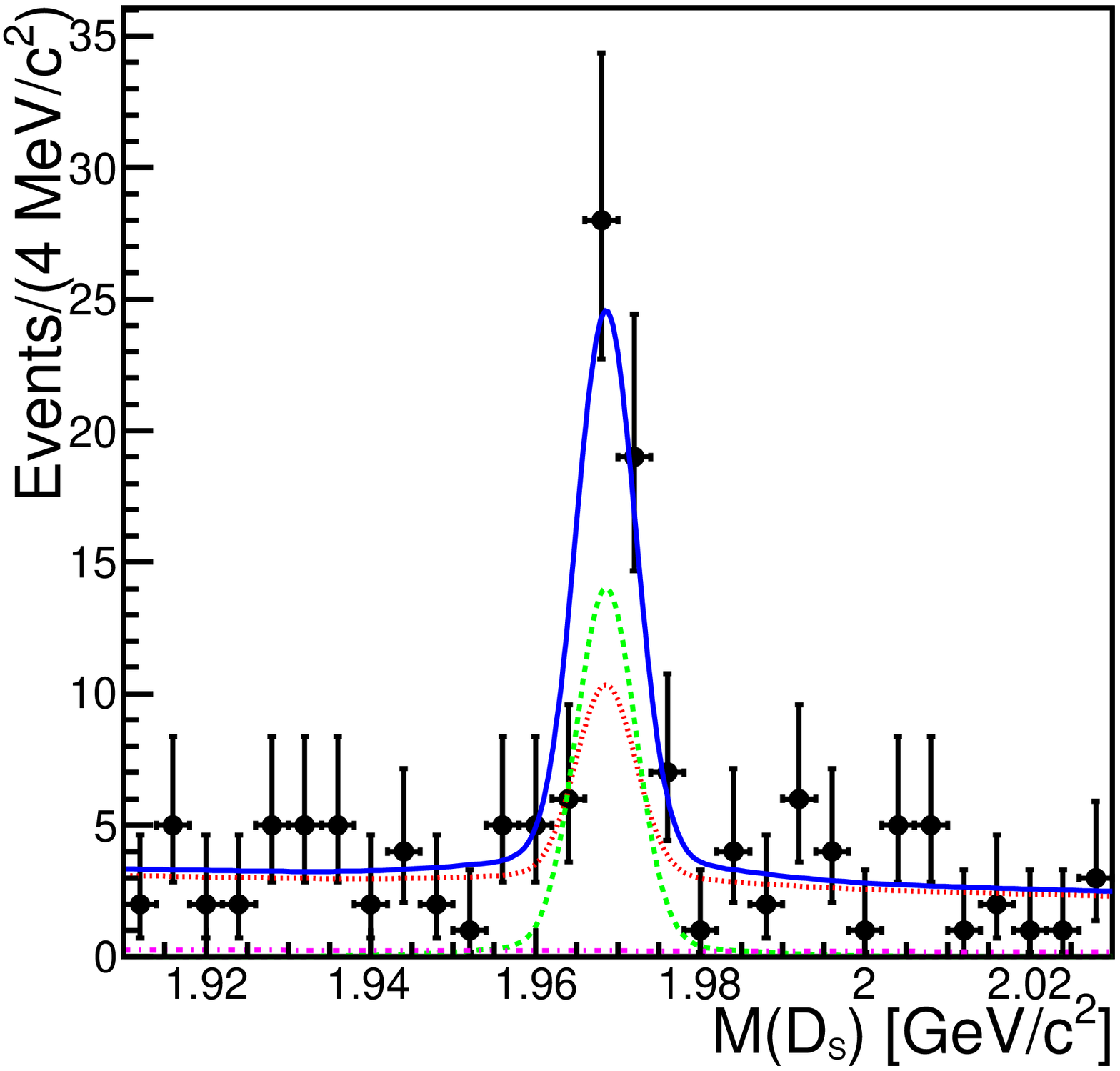}
\end{picture}
\end{minipage}\hfill
\begin{minipage}[b]{.32\linewidth}
\centering
\setlength{\unitlength}{1mm}
\begin{picture}(60,50)
\includegraphics[height=4.9cm,width=5.9cm]{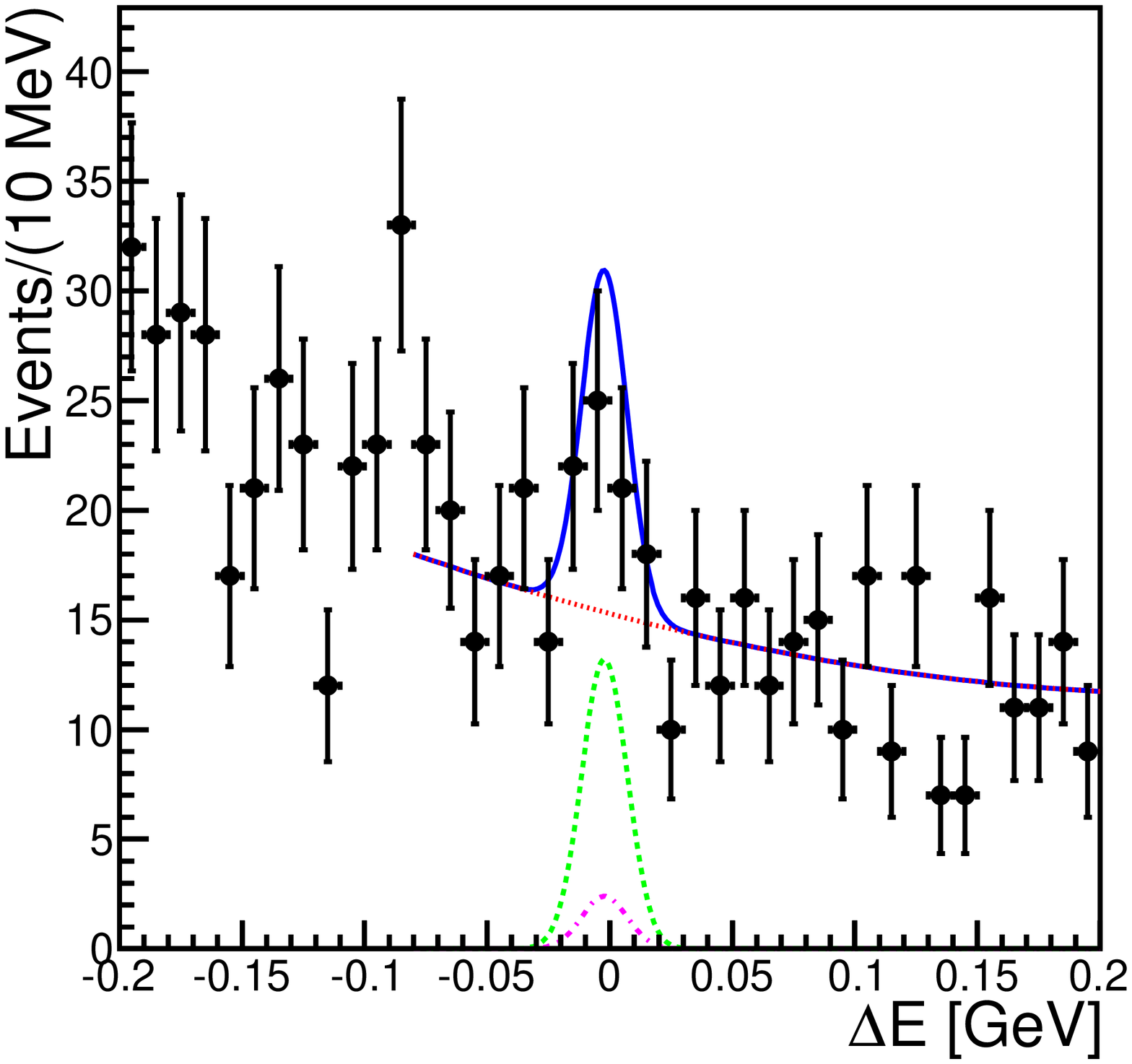}
\end{picture}
\end{minipage}\hfill
\begin{minipage}[b]{.32\linewidth}
\centering
\setlength{\unitlength}{1mm}
\begin{picture}(60,50)
\includegraphics[height=4.9cm,width=5.9cm]{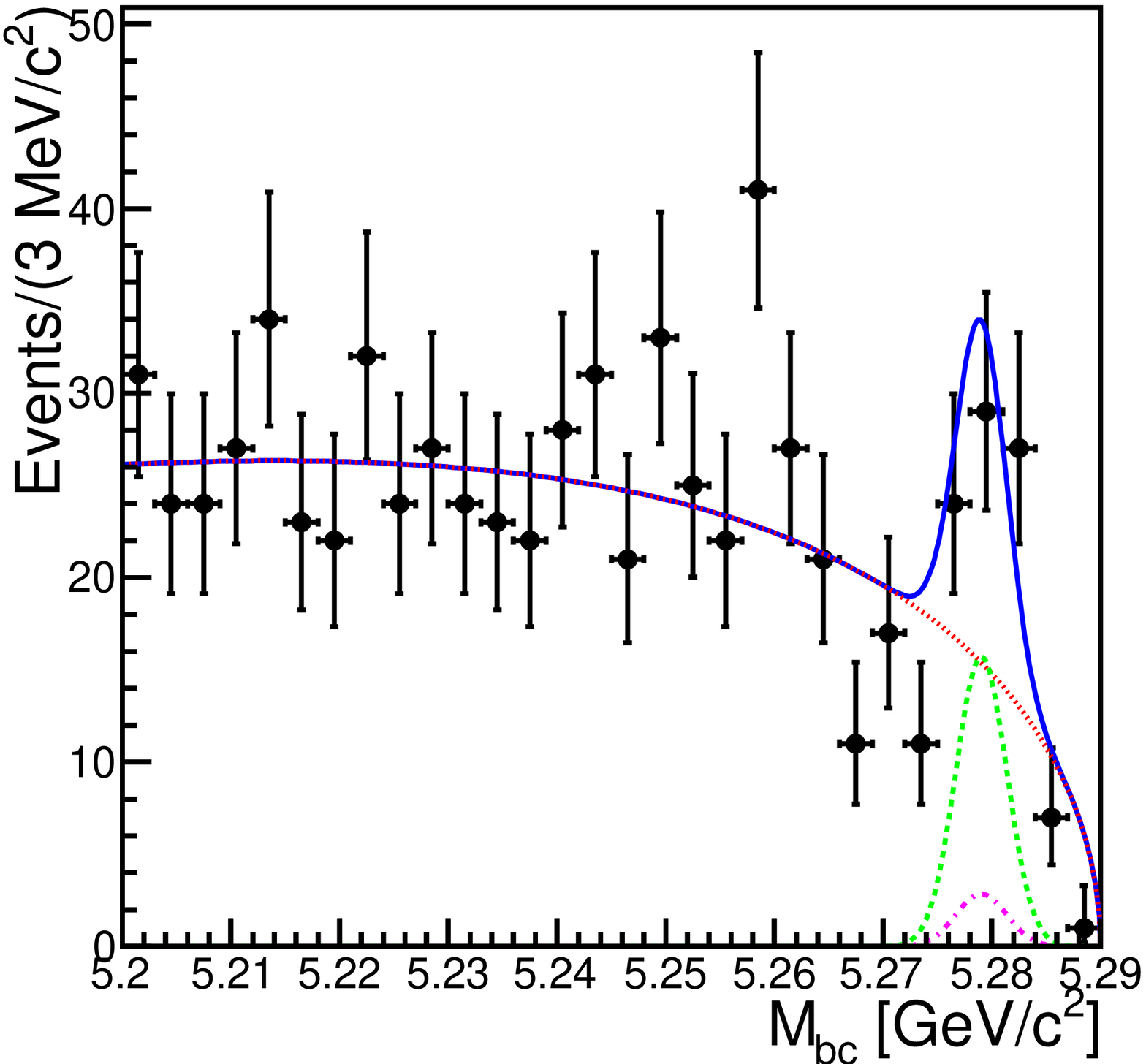}
\end{picture}
\end{minipage}\hfill
\begin{minipage}[b]{.32\linewidth}
\centering
\setlength{\unitlength}{1mm}
\begin{picture}(60,50)
\includegraphics[height=4.9cm,width=5.9cm]{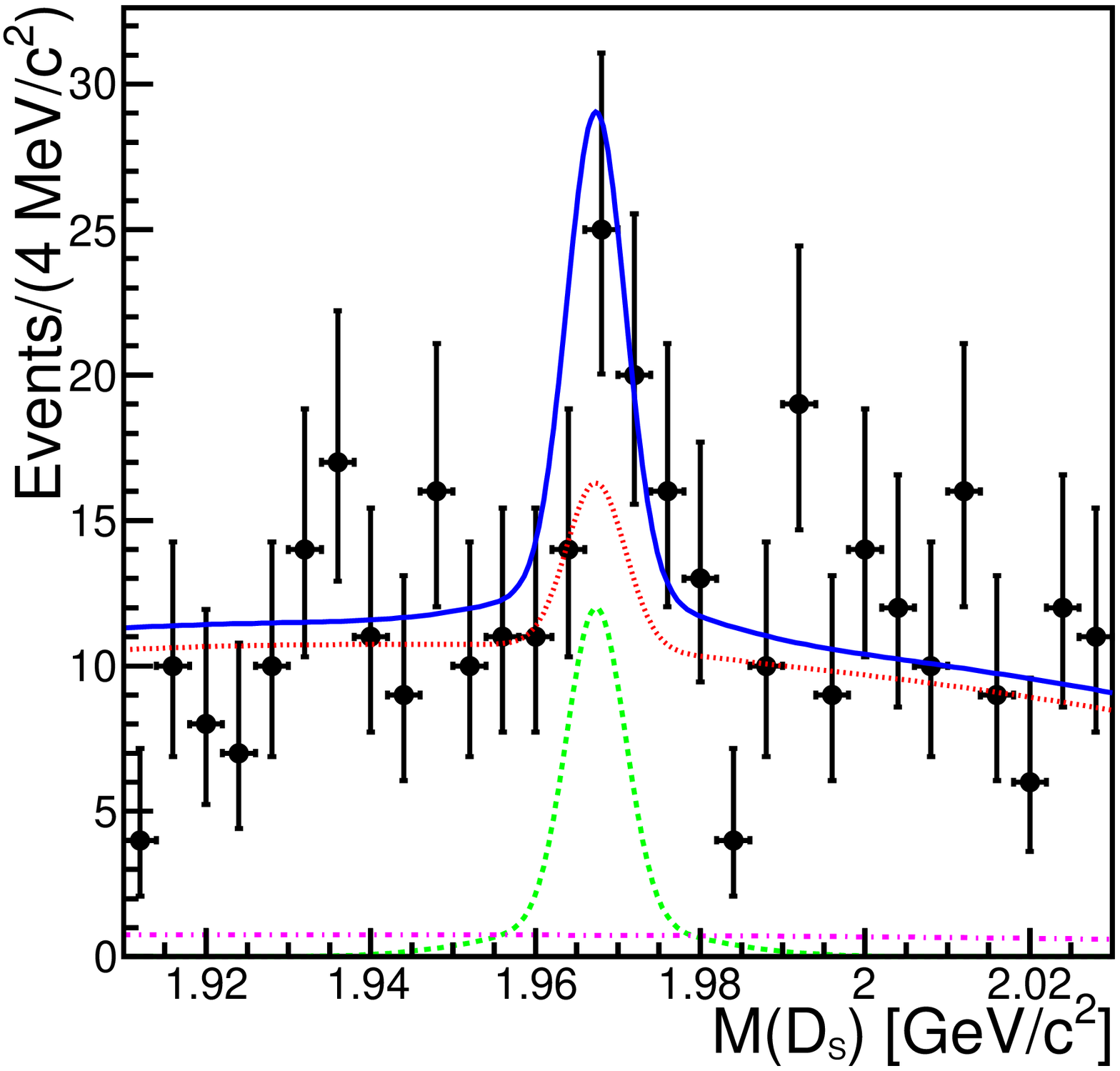}
\end{picture}
\end{minipage}\hfill
\begin{minipage}[b]{.32\linewidth}
\centering
\setlength{\unitlength}{1mm}
\begin{picture}(60,50)
\includegraphics[height=4.9cm,width=5.9cm]{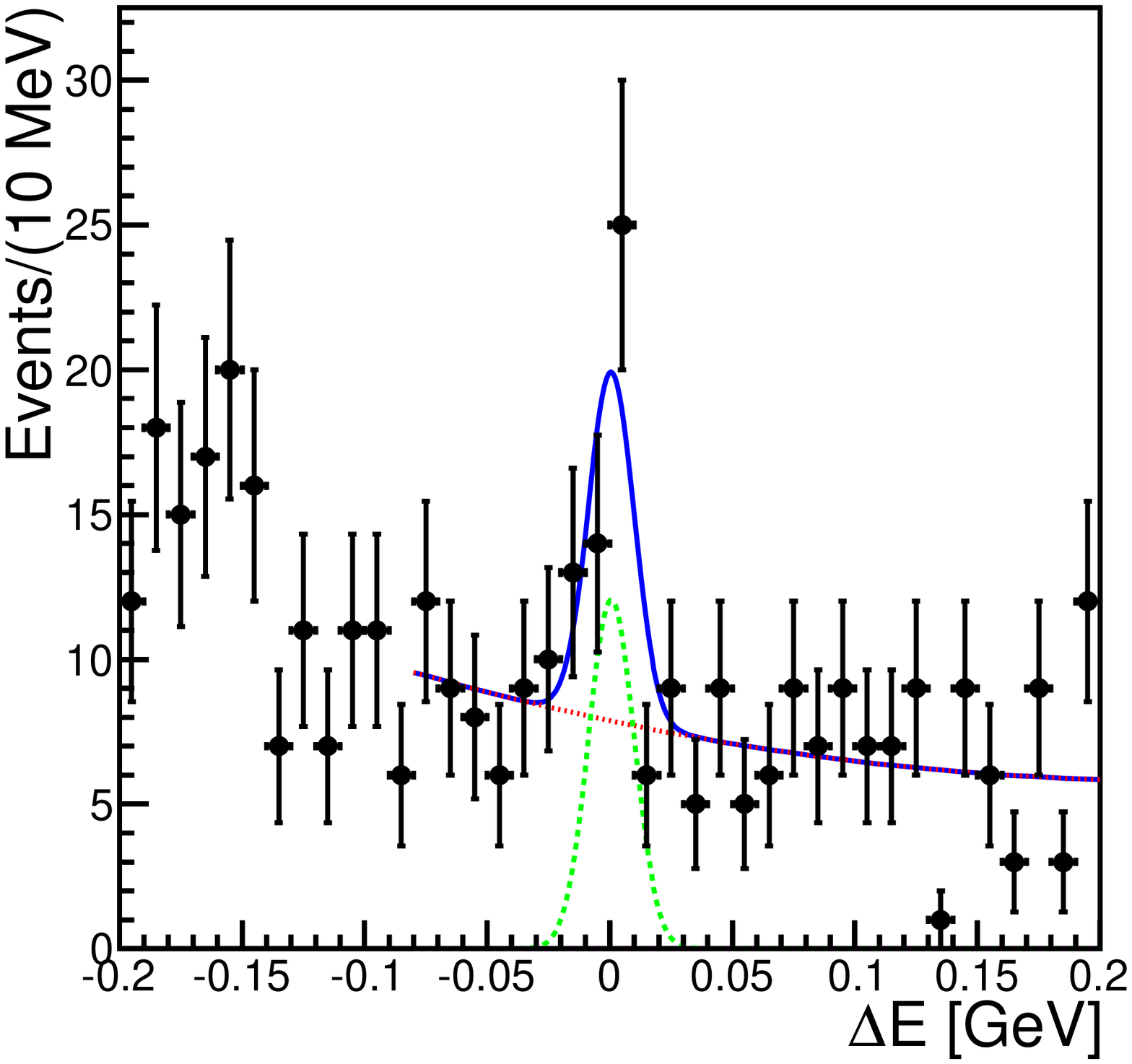}
\end{picture}
\end{minipage}\hfill
\begin{minipage}[b]{.32\linewidth}
\centering
\setlength{\unitlength}{1mm}
\begin{picture}(60,50)
\includegraphics[height=4.9cm,width=5.9cm]{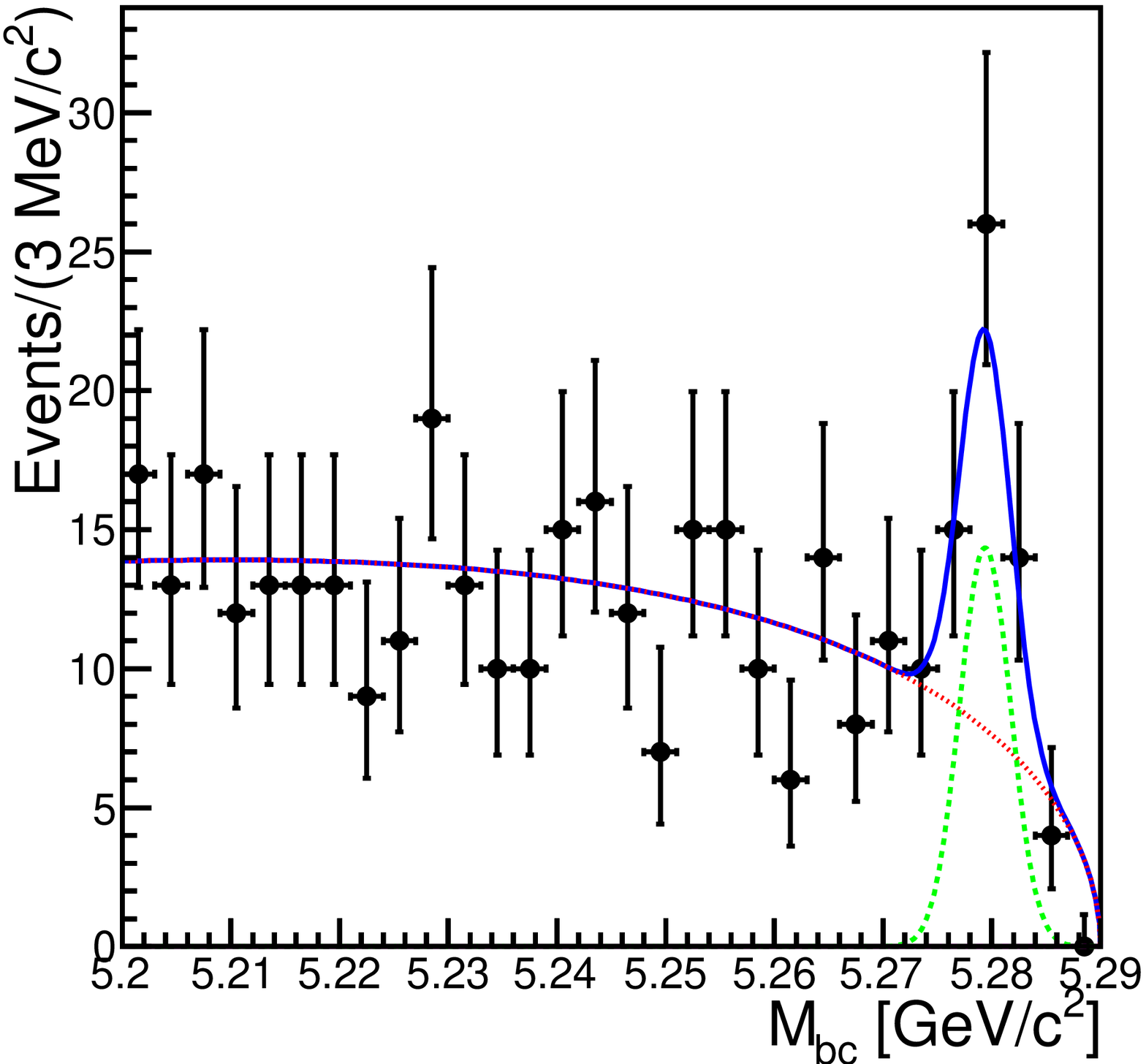}
\end{picture}
\end{minipage}\hfill
\begin{minipage}[b]{.32\linewidth}
\centering
\setlength{\unitlength}{1mm}
\begin{picture}(60,50)
\includegraphics[height=4.9cm,width=5.9cm]{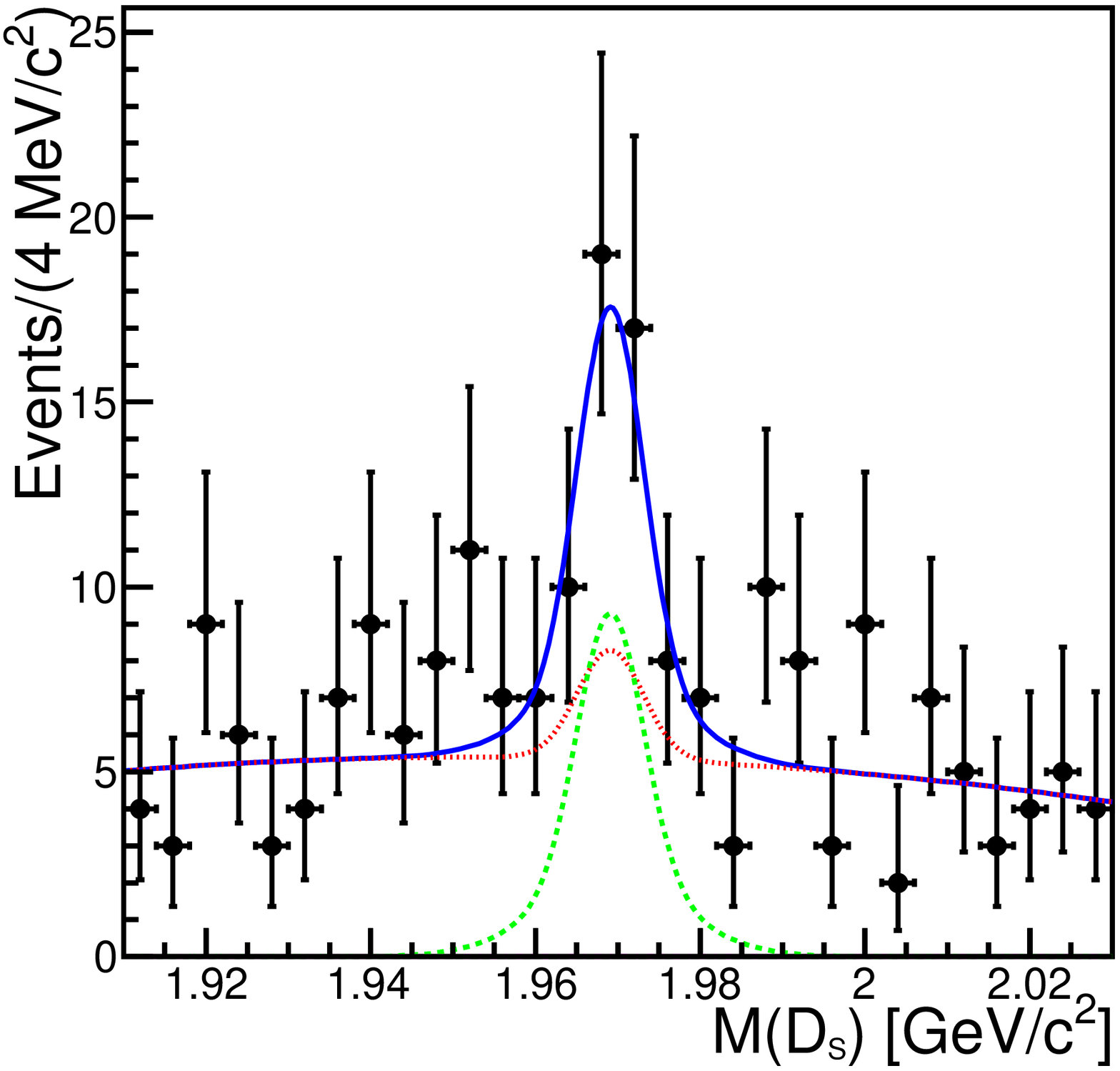}
\end{picture}
\end{minipage}\hfill
\caption{Distributions of $\Delta E$,  $M_{\rm bc}$ and $M(D_s)$ for (top) $B^0\to D^{-}_s(\to \phi\pi^-) K^0_S\pi^+$,  (middle) $B^0\to D^{-}_s(\to K^{*0} K^-) K^0_S\pi^+$, and (bottom) $B^0\to D^{-}_s(\to K^0_S K^-) K^0_S\pi^+$ decays. The distribution for each quantity is shown in the signal region of the remaining two.
The blue solid curves show the results of the overall fit described in the text, the green dotted curves correspond to the signal component, the red long-dashed curves indicate the combinatorial background (including the peaking $D_s$ component) and the pink dot-dashed curves represent the peaking $B^0$ background.}
\label{FIG_DSKSPI}
\end{figure*}

A $B$ candidate is reconstructed by combining the $D_s$ candidate with a selected $K^0_S$ and a charged pion for $B^0 \to D_s^-K^0_S \pi^+$, and with a pair of kaons of the same charge for $B^+ \to D_s^-K^+K^+$. A quality requirement on the $B$ vertex-fit statistic ($\chi_B^2/{\rm NDF} < 60$) to the $D_s^-K^+K^+$ ($D_s^-K^0_S \pi^+$) trajectories is applied, where the $D_s$ mass is constrained to its world-average value~\cite{PDG} and NDF is
the number of degrees of freedom. The signal decays are  identified 
by three kinematic variables: the $D_s$ invariant mass, the energy difference
  $\Delta E = E_B - E_\mathrm{beam}$, and the beam-energy-constrained mass
 $ M_{\rm bc} = (\sqrt{E_\mathrm{beam}^2 - |\vec{p}_B|^2c^2})/c^2$.
Here, $E_B$ and $\vec{p}_B$ are the reconstructed energy and momentum of the
$B$ candidate, respectively, and $E_\mathrm{beam}$ is the run-dependent beam energy,
all calculated in the $e^+e^-$ center-of-mass (CM) frame.
We retain candidate events in the three-dimensional region defined by
 $1.91~{\rm GeV/}c^2 < M(D_s) < 2.03$ GeV/$c^2$,
 $5.2~{\rm GeV/}c^2 < M_{\rm bc} < 5.3$ GeV/$c^2$
and $ -0.2~{\rm GeV} < \Delta E < 0.2$ GeV. In the fit described later, we use a narrower range $ -0.08~{\rm GeV} < \Delta E < 0.20~{\rm GeV}$ to exclude the possible contamination from
$B\to D_s X$ decays having higher multiplicities.
From a GEANT3~\cite{GEANT} based Monte Carlo (MC) simulation, we find 
the signal peaks in a region defined by
 $1.9532~{\rm GeV/}c^2 < M(D_s) < 1.9832$ GeV/$ c^2$,
 $5.27~{\rm GeV/}c^2 < M_{\rm bc} < 5.29$ GeV/$c^2$ and
$|\Delta E| < 0.03$ GeV.
Based on MC simulation, the region  2.88 GeV/$c^2 <
M(c\overline{c}) <$ 3.18 GeV/$c^2$ is excluded
to remove background from $B^+\to (c\overline{c}) K^+$ or $B^0\to (c\overline{c}) K^0_S$ decays, where
$(c\overline{c})$ denotes a charmonium state such as the $J/\psi$ or $\eta_c$ and $M(c\overline{c})$ is the invariant mass of its decay products ($K^+K^-\pi^+\pi^-$ or $K^0_S K^+\pi^-$ for the corresponding $D_s$ mode).

We find that, for the $B^0 \to D_s^-K^0_S \pi^+$ ($B^+ \to D_s^-K^+K^+$) decays, the average number of $B$ candidates per event is 1.14 (1.04). If an event has more than one $B$ candidate, we select the one
with the smallest value of $\chi_B^2$.

We exploit the event topology to discriminate between spherical
$B\overline{B}$ events and the dominant background from jet-like continuum
$e^{+}e^{-} \to q\overline{q}$ ($q$ = $u$, $d$, $s$,
$c$) events. We require the event shape variable $R_2$, defined as the ratio
of the second- and zeroth-order Fox-Wolfram moments~\cite{FOX}, 
to be less than 0.4 to suppress the continuum background. 

%%%%%%%%%%%%%%%%%%%%%%%%%%%%%%%%%%%%%%%%%%%%%%%%%%%%%%%%%%%%%%%%%%%%%%%%%%%%%%%%%%%%%%%%%%%%%%%%%%
%%%%%%%%%%%%%%%%%%%%%%%%%%%%%%%%%%%%%%%%%%%%%%%%%%%%%%%%%%%%%%%%%%%%%%%%%%%%%%%%%%%%%%%%%%%%%%%%%%
%%%%%%%%%%%%%%%%%%%%%%%%%%%%%%%%%%%%%%%%%%%%%%%%%%%%%%%%%%%%%%%%%%%%%%%%%%%%%%%%%%%%%%%%%%%%%%%%%%

%\begin{widetext}

\begin{figure*}[htb]
\begin{minipage}[b]{.32\linewidth}
\centering
\setlength{\unitlength}{1mm}
\begin{picture}(60,50)
\includegraphics[height=4.9cm,width=5.9cm]{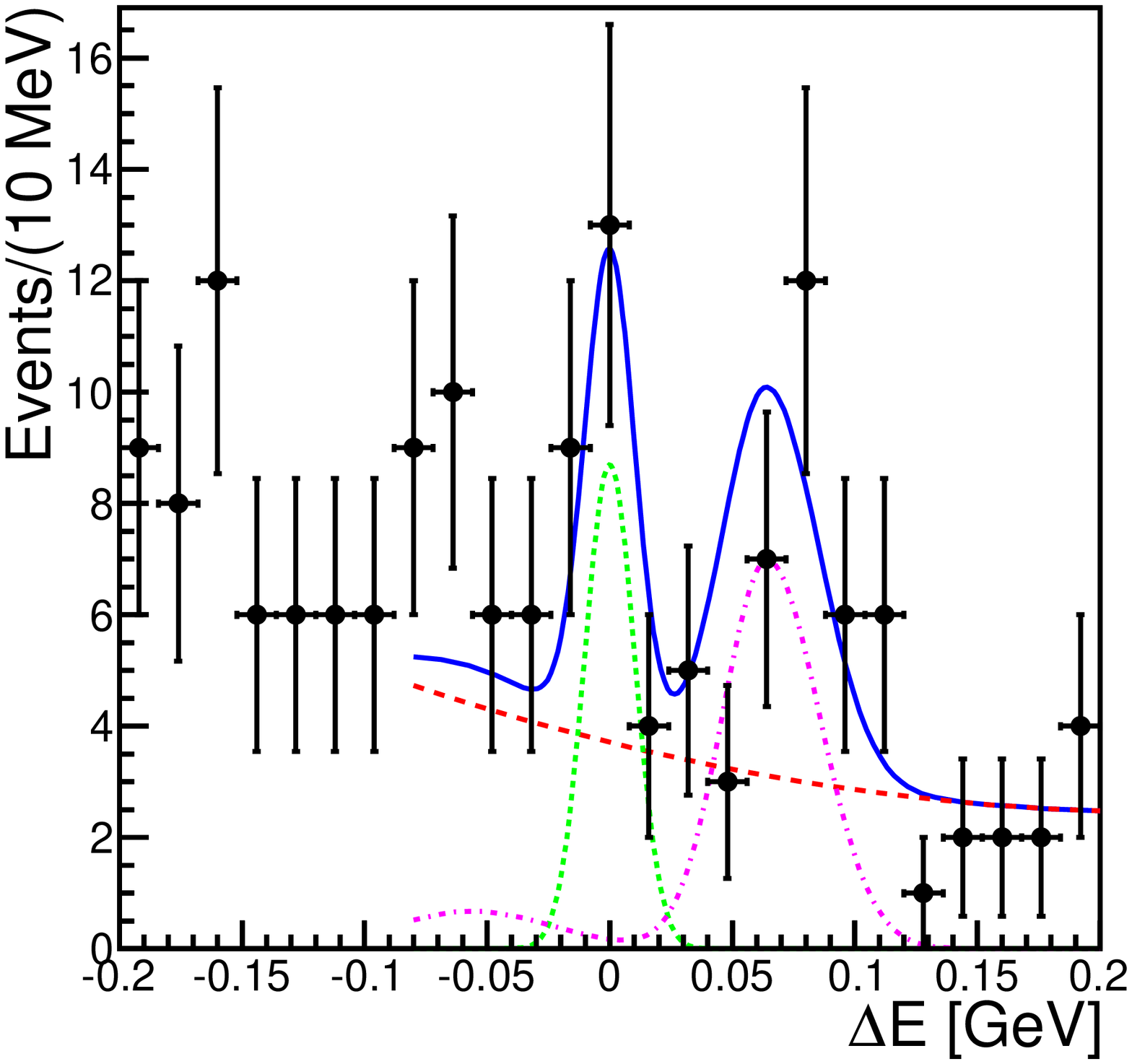}
\end{picture}
\end{minipage}\hfill
\begin{minipage}[b]{.32\linewidth}
\centering
\setlength{\unitlength}{1mm}
\begin{picture}(60,50)
\includegraphics[height=4.9cm,width=5.9cm]{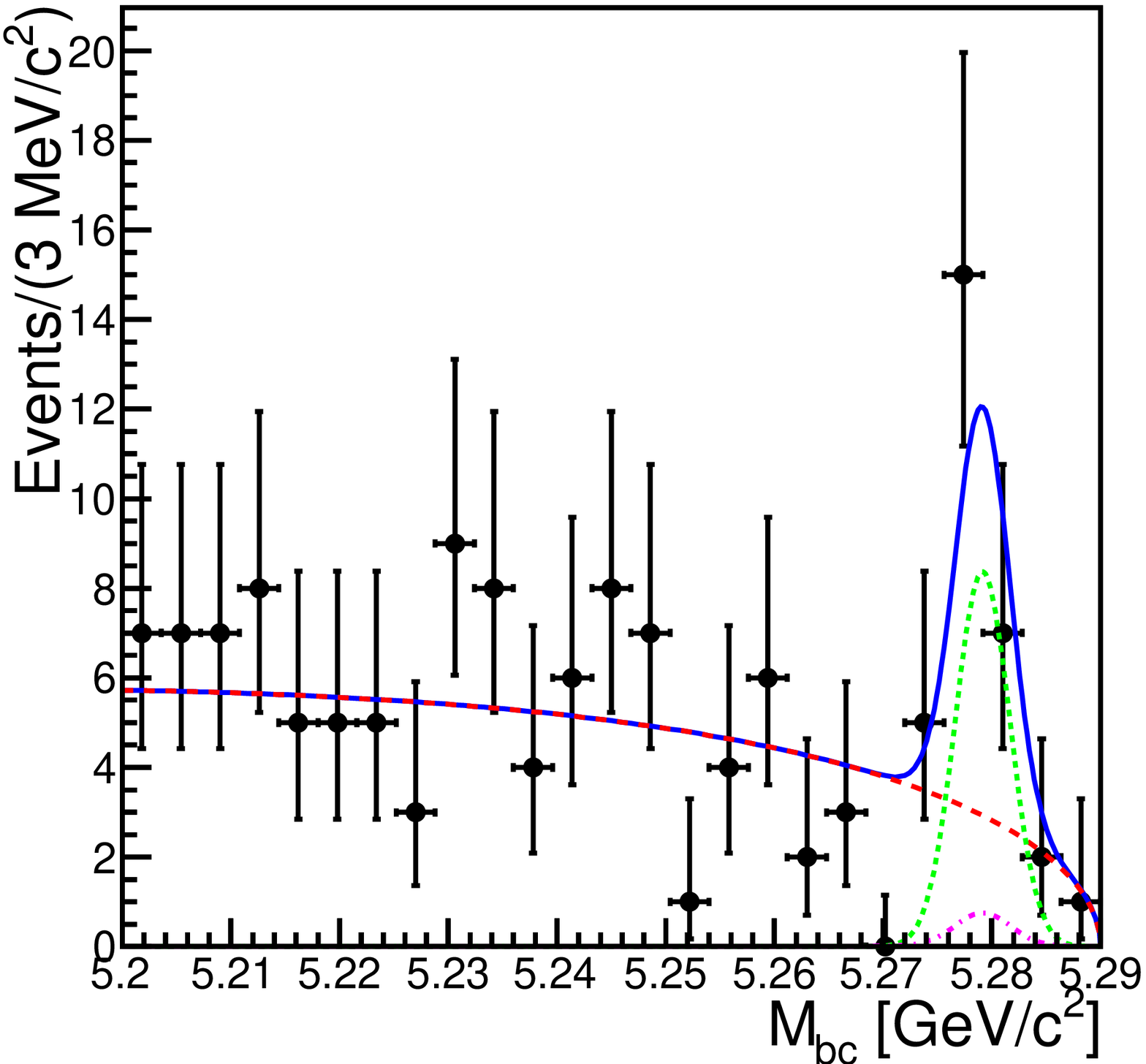}
\end{picture}
\end{minipage}\hfill
\begin{minipage}[b]{.32\linewidth}
\centering
\setlength{\unitlength}{1mm}
\begin{picture}(60,50)
\includegraphics[height=4.9cm,width=5.9cm]{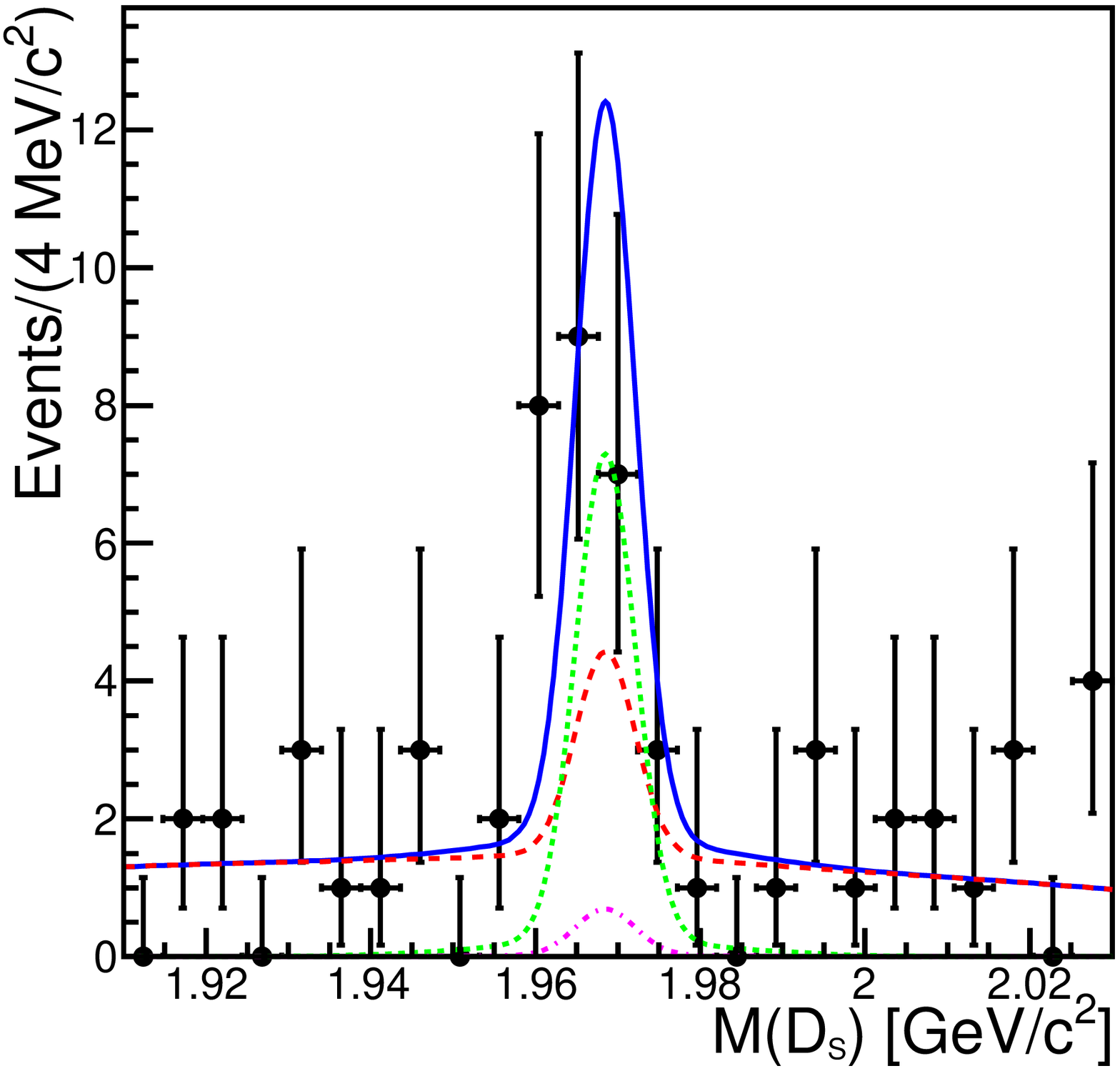}
\end{picture}
\end{minipage}\hfill
\begin{minipage}[b]{.32\linewidth}
\centering
\setlength{\unitlength}{1mm}
\begin{picture}(60,50)
\includegraphics[height=4.9cm,width=5.9cm]{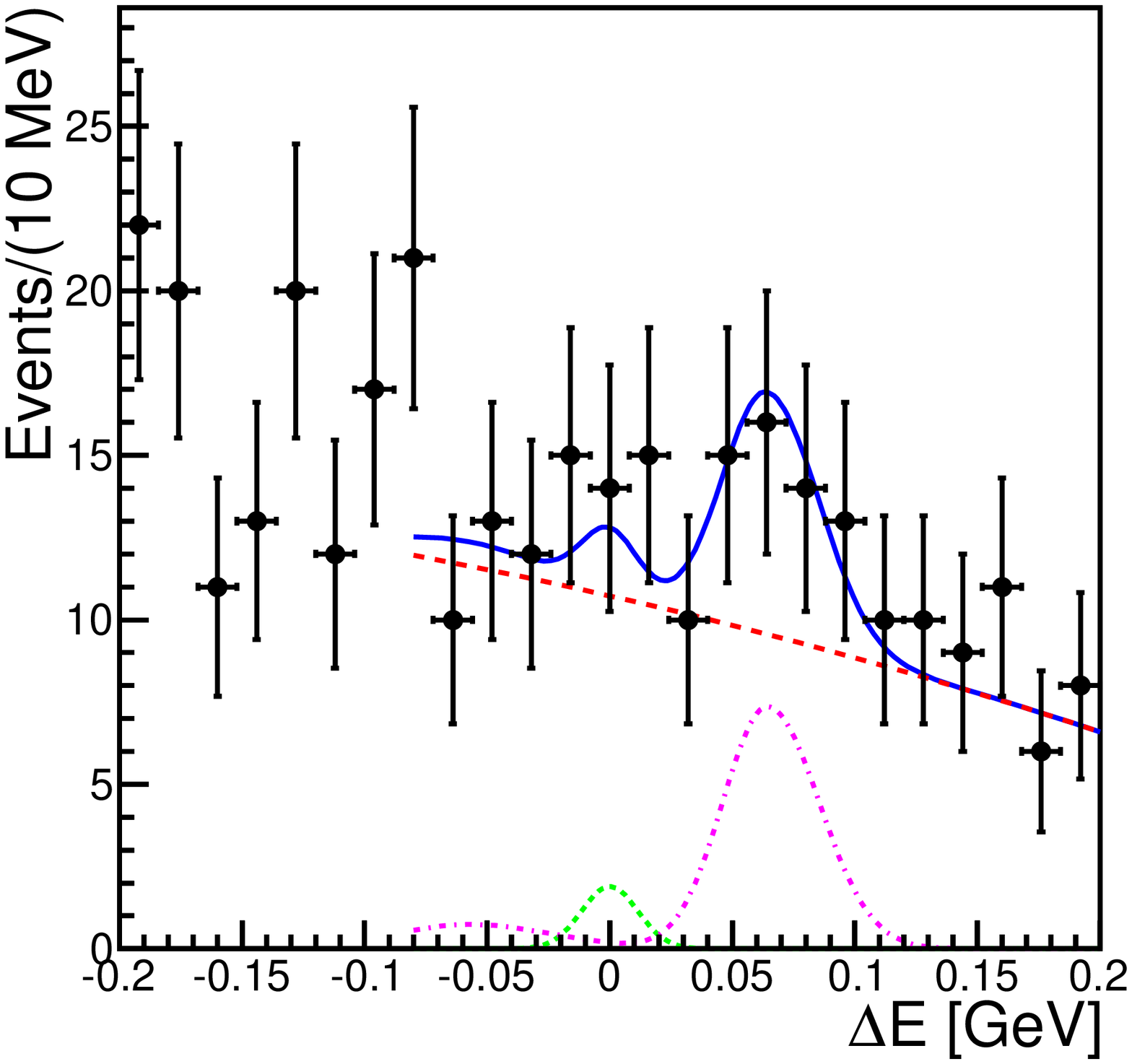}
\end{picture}
\end{minipage}\hfill
\begin{minipage}[b]{.32\linewidth}
\centering
\setlength{\unitlength}{1mm}
\begin{picture}(60,50)
\includegraphics[height=4.9cm,width=5.9cm]{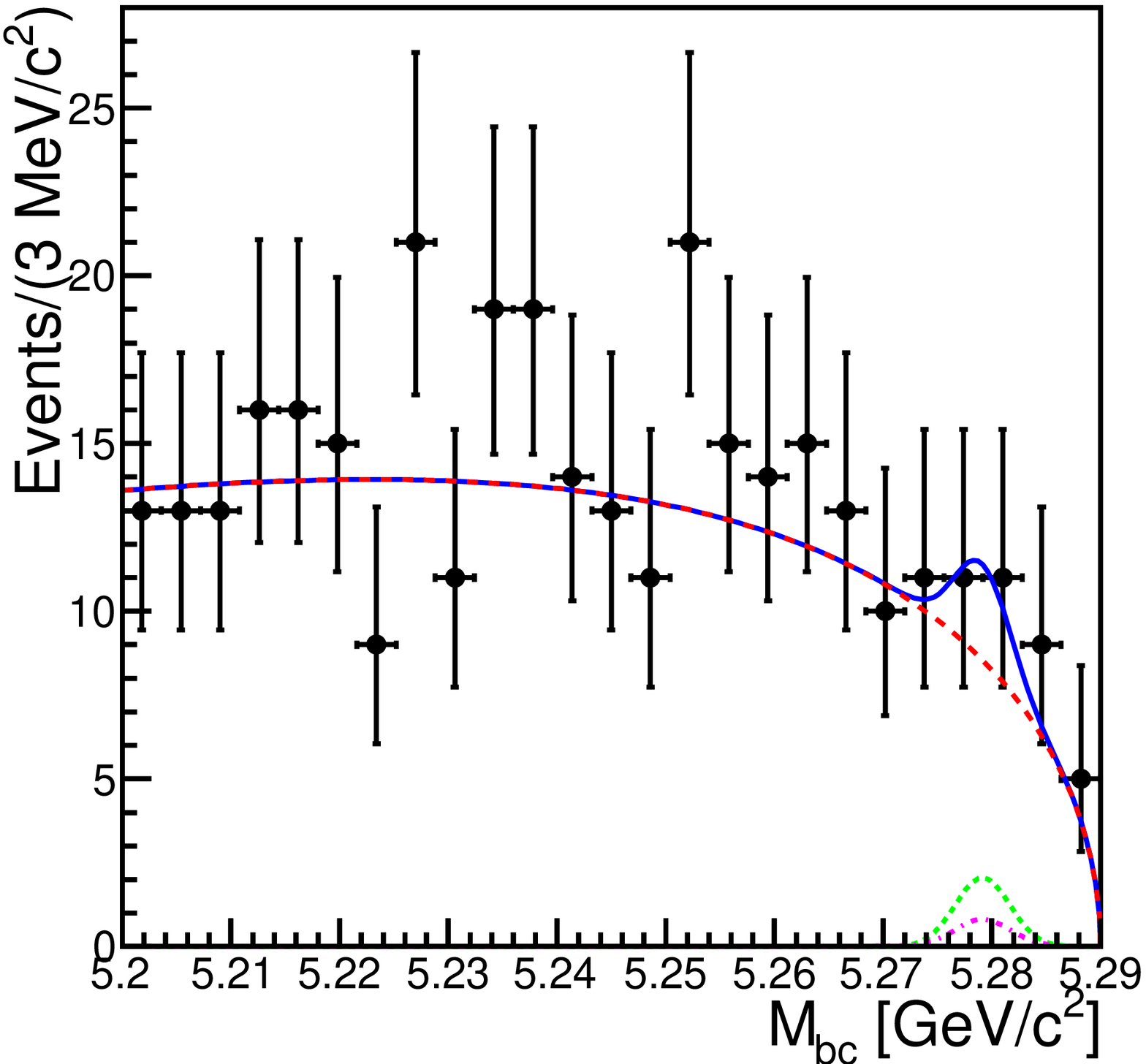}
\end{picture}
\end{minipage}\hfill
\begin{minipage}[b]{.32\linewidth}
\centering
\setlength{\unitlength}{1mm}
\begin{picture}(60,50)
\includegraphics[height=4.9cm,width=5.9cm]{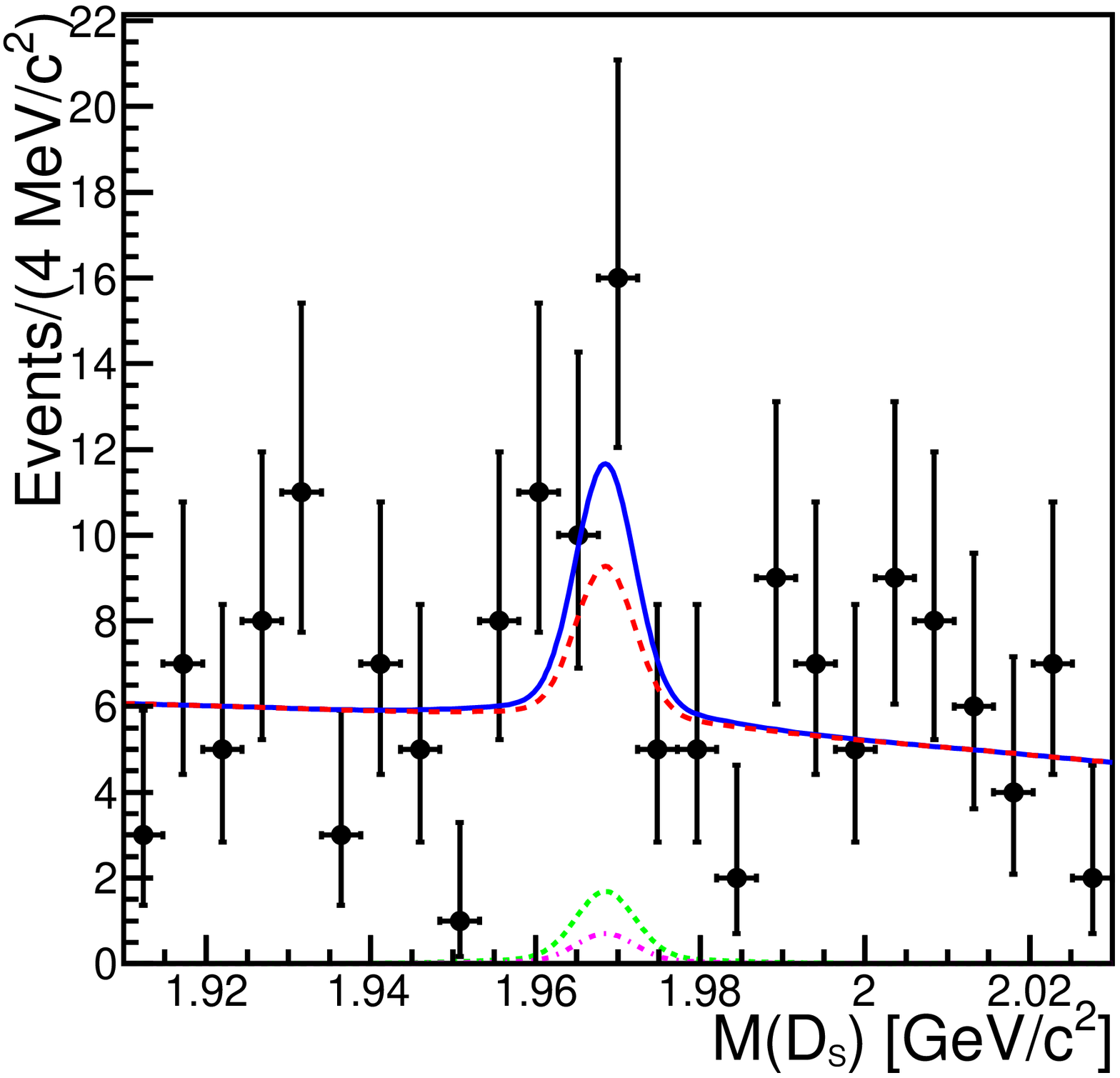}
\end{picture}
\end{minipage}\hfill
\begin{minipage}[b]{.32\linewidth}
\centering
\setlength{\unitlength}{1mm}
\begin{picture}(60,50)
\includegraphics[height=4.9cm,width=5.9cm]{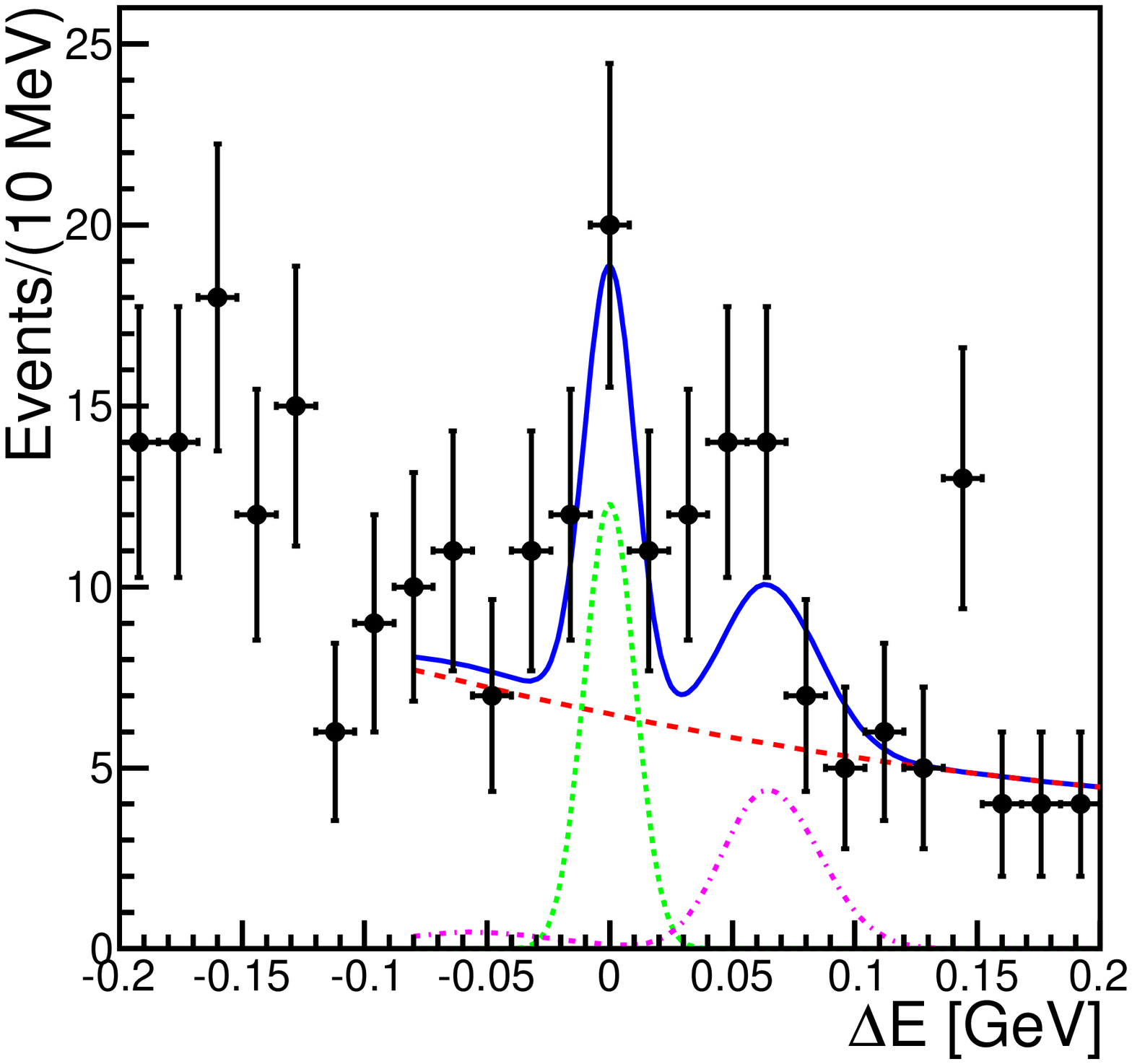}
\end{picture}
\end{minipage}\hfill
\begin{minipage}[b]{.32\linewidth}
\centering
\setlength{\unitlength}{1mm}
\begin{picture}(60,50)
\includegraphics[height=4.9cm,width=5.9cm]{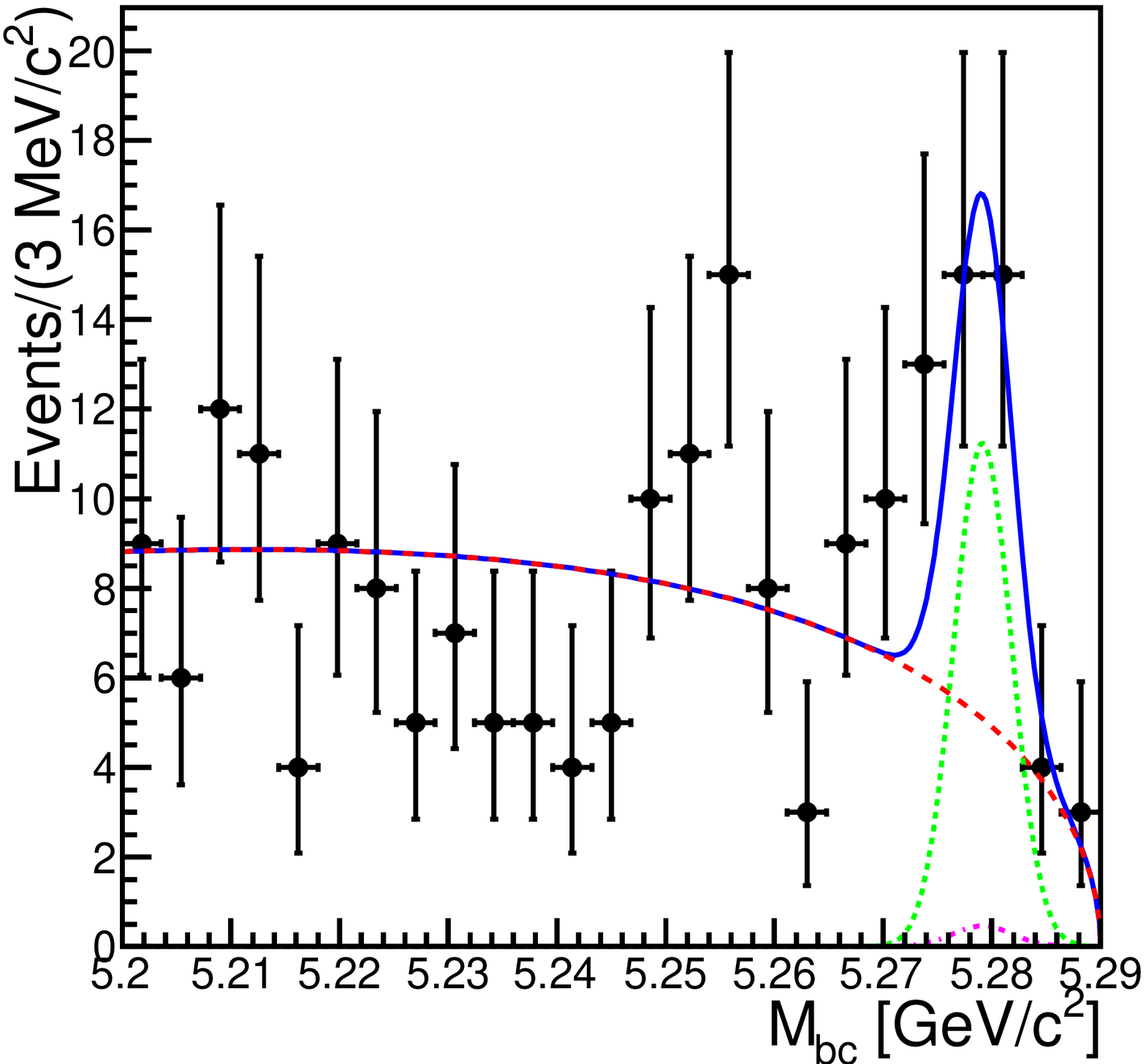}
\end{picture}
\end{minipage}\hfill
\begin{minipage}[b]{.32\linewidth}
\centering
\setlength{\unitlength}{1mm}
\begin{picture}(60,50)
\includegraphics[height=4.9cm,width=5.9cm]{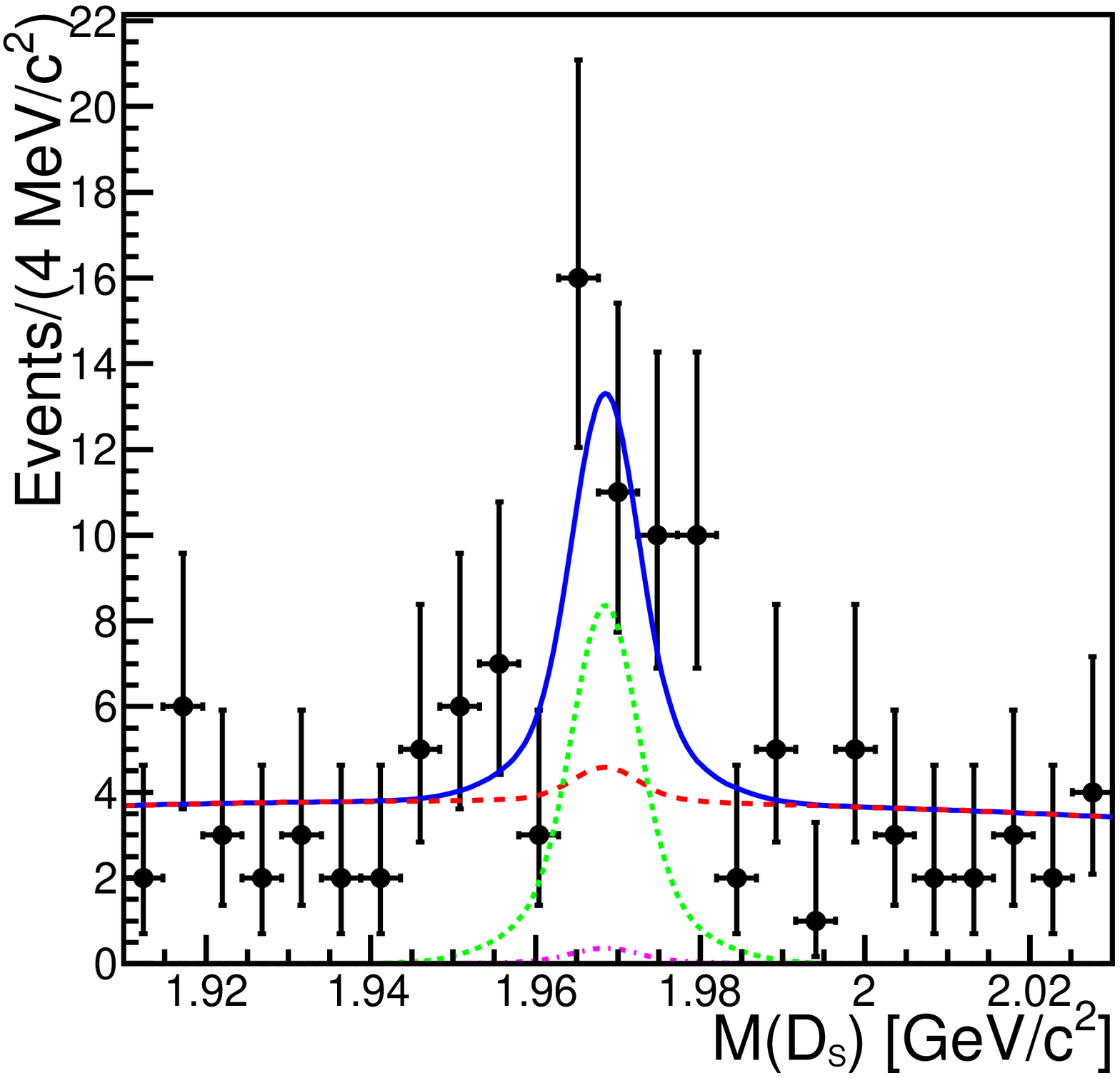}
\end{picture}
\end{minipage}\hfill
\caption{Distributions of $\Delta E$,  $M_{\rm bc}$ and $M(D_s)$
for (top)  $B^+\to D^{-}_s(\to \phi\pi^-) K^+K^+$,
(middle) $B^+\to D^{-}_s(\to K^{*0} K^-) K^+K^+$, and 
(bottom) $B^+\to D^{-}_s(\to K^0_S K^-) K^+K^+$ decays.
The distribution for each quantity is shown in the signal region of the remaining two.
The blue solid curves show the results of the overall fit described in the text, the green dotted curves correspond to the signal component, the red long-dashed curves indicate the combinatorial background (including the peaking $D_s$ component) and the pink dot-dashed curves represent the $B \to D_s^{(*)} K\pi$ contribution.}
\label{FIG_DSKK}
\end{figure*}

%%%%%%%%%%%%%%%%%%%%%%%%%%%%%%%%%%%%%%%%%%%%%%%%%%%%%%%%%%%%%%%%%%%%%%%%%%%%%%%%%%%%%%%%%%%%%%%%%%
%%%%%%%%%%%%%%%%%%%%%%%%%%%%%%%%%%%%%%%%%%%%%%%%%%%%%%%%%%%%%%%%%%%%%%%%%%%%%%%%%%%%%%%%%%%%%%%%%%

Large MC samples are used to evaluate possible background from $B\bar{B}$ and continuum $q\bar{q}$ events for both studied channels. In the $B^0 \to D_s^-K^0_S \pi^+$ analysis, a significant contribution from $B^0 \to D_s^-D^+$, $D^+ \to K^0_S \pi^+$ is identified. We require the quantity $|M(K^0_S \pi^+) - m_{D^+}|$ to be less (greater) than  30 MeV/$c^2$ to select the $B^0 \to D_s^-D^+$ control sample (to suppress the charm contribution), where $m_{D^+}$ is the world-average value of the $D^+$ meson mass. Furthermore, we identify a peaking background arising from the $B^0$ decaying to the same final state of five hadrons (``$B^0$ peaking background''). Such events do not contain a $D_s$ meson in the decay chain and mainly include ($c\bar{c}$) states like $\psi(2S)$, $\eta_c(2S)$, $\chi_{c1}(1P)$ and $\chi_{c0}(1P)$. We find a significant contribution to $B^+ \to D_s^-K^+K^+$ from the $B^+ \to D_s^{(*)-} K^+\pi^+$ decays owing to pion misidentification (or a missing photon in the $D_s^*$ reconstruction).
We evaluate the shape of this contribution in the $\Delta E$,  $M_{\rm bc}$ and $M(D_s)$ projections using MC samples of the $B^+ \to D_s^{(*)-} K^+\pi^+$ processes after subjecting them to the $B^+ \to D_s^-K^+K^+$ selection.
Finally, we identify an additional background contribution containing good $D_s$ candidates randomly combined with $K^+K^+$ or $K^0_S\pi^+$ (``$D_s$ peaking background'').
All aforementioned background contributions are taken into account in our fitting procedures.

The signal yields are obtained from unbinned extended maximum-likelihood fits
to the  [$\Delta E, M_{\rm bc},
M(D_s)$] distributions of the selected candidate events.
The likelihood function is given by
\begin{figure*}[t]
\includegraphics[height=8cm,width=16cm]{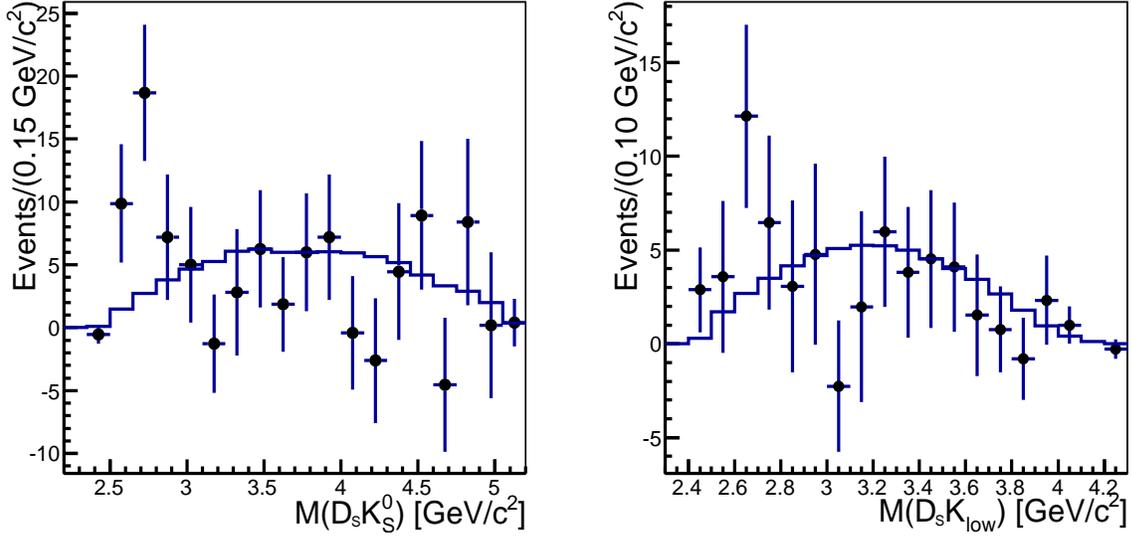}
\caption{ Invariant mass distributions of (left) $D_s^- K^0_S$ for the $B^0 \to D_s^- K^0_S \pi^+$  and (right) $D_s^- K^+_{\rm low}$ for $B^+ \to D_s^-K^+K^+$ decay events in the signal region of $\Delta E$, $M_{\rm bc}$ and $M_{D_s}$ after applying all selection criteria. Points with error bars represent the data after subtraction of the background contribution, estimated from the $M_{\rm bc}$ sideband ($5.22~{\rm GeV}/c^2 < M_{\rm bc} < 5.26~{\rm GeV}/c^2$). The histograms show the phase-space distribution of the signal MC sample normalized to the data luminosity.
 }
\label{DSKPAIR}
\end{figure*} 
\begin{equation}
{\cal L} = \frac{1}{N!} \cdot \exp\Big(-\sum_{j} N_j\Big) 
\cdot \prod_{i=1}^{N} \Big( \sum_{j} N_j{\cal P}_i^j \Big)\, ,
\end{equation}
% czerwone
where $j$ runs over the signal and background components, $i$ is the event index, $N_j$ and ${\cal P}_i^j$ denote the yield and probability density functions (PDFs) for each component, respectively, and $N$ is the total number of data events.
Neglecting the small correlation between each pair of fit observables, we construct the overall PDF as a product of their individual PDFs.
Two components, signal and combinatorial background ($j=$ sig, cmb), are common for $B^0 \to D_s^-K^0_S \pi^+$ and
$B^+ \to D_s^-K^+K^+$. Their respective PDF parameterizations are constructed as
\begin{equation}
{\cal P}^{\rm sig}_i = {\cal G}(\Delta E^i;~\overline{\Delta E}, \sigma_{\Delta E}) \times {\cal G}(M^i_{\rm bc};~ m_B, \sigma_{M_{\rm bc}}) \times {\cal G}_2^{\rm sig}\bigl(M^i(D_s);~m_{D_s}, \sigma^{(1)}_{D_s}, \sigma^{(2)}_{D_s}, f^{\rm sig}_{D_s}\bigr)
\label{EQ1}
\end{equation}
and
\begin{equation}
\begin{split}
{\cal P}^{{\rm cmb}}_i &=  p_2(\Delta E^i;~w_0,w_1,w_2)\times A(M^i_{\rm bc};~\zeta)
 \times \Big[f^{{\rm peak}}_{D_s} \cdot {\cal G}_2^{\rm bkg}\bigl(M^i(D_s);~m_{D_s}, \sigma^{(1)}_{D_s}, \sigma^{(2)}_{D_s}, f^{\rm bkg}_{D_s}\bigr)\\
&+ (1-f^{{\rm peak}}_{D_s}) \cdot p_2(M^i(D_s);~v_0,v_1,v_2)\Big].
\end{split}
\label{EQ2}
\end{equation}
Here, we use a Gaussian function (${\cal G}$) to parameterize the signal PDF in $\Delta E$ and $M_{\rm bc}$
and a double-Gaussian function (${\cal G}_2$) with a common mean for the $M(D_s)$ distribution.
The combinatorial background component utilizes a second-order Chebyshev polynomial ($p_2$) in the $\Delta E$ distribution and an ARGUS function~\cite{ARGUS}, $A(M_{\rm bc},\zeta) \propto M_{\rm bc} \cdot \sqrt{1-(M_{\rm bc}/E_\mathrm{beam})^2} \cdot e^{-\zeta(1-(M_{\rm bc}/E_\mathrm{beam})^2)}$ for the $M_{\rm{bc}}$ distribution, where $\zeta$ is a fit parameter. 
The combinatorial background's $M(D_s)$ distribution is described by the sum of a double-Gaussian function for the ``$D_s$ peaking background'' and a second-order Chebyshev polynomial with a relative fraction $f^{{\rm peak}}_{D_s}$ of these two components. The double-Gaussian function for component $j$ is defined as

\begin{equation}
\begin{split}
{\cal G}_2^j\bigl(M^i(D_s);~m_{D_s}, \sigma^{(1)}_{D_s}, \sigma^{(2)}_{D_s}, f^j_{D_s}\bigr) &= f^j_{D_s} \cdot {\cal G}\bigl(M^i(D_s);~m_{D_s}, \sigma^{(1)}_{D_s}\bigr) \\
&+ (1-f^j_{D_s}) \cdot {\cal G}\bigl(M^i(D_s);~m_{D_s}, \sigma^{(2)}_{D_s}\bigr),
\end{split}
\label{EQ3}
\end{equation}
where $f^j_{D_s}$ denotes the relative contribution of the core over the tail Gaussian in the $M(D_s)$ distribution.

In Eqs.~(\ref{EQ1}-\ref{EQ3}), $\overline{\Delta E}, m_B, m_{D_s}, \sigma_{\Delta E}, \sigma_{M_{\rm bc}}, \sigma^{(1)}_{D_s}$, $\sigma^{(2)}_{D_s}$ (the respective mean values and widths of the Gaussians), $f^{{\rm peak}}_{D_s}$ and $f^{{\rm sig}({\rm bkg})}_{D_s}$ are fit parameters.
For both channels studied, the parameters $\sigma^{(1)}_{D_s}$, $\sigma^{(2)}_{D_s}$ and $f^{{\rm sig}({\rm bkg})}_{D_s}$ are fixed to the values obtained from the $B^+\to D_s^+\overline{D}{}^0$ control channel. In addition, we use the $B^0 \to D_s^-D^+$ ($B^+ \to D_s^+\overline{D}{}^0$) control sample to determine the signal width values for the $\Delta E$ and $M_{\rm bc}$ distributions that are later fixed in the fit to the $B^0 \to D_s^-K^0_S \pi^+$ ($B^+ \to D_s^-K^+K^+$) data sample. 

\begin{table*}[t]
\begin{center}
\caption{Signal yields, average reconstruction efficiencies, statistical significances and branching fractions for 
$B^0 \to D_s^-K^0_S \pi^+$ and $B^+ \to D_s^-K^+K^+$ decays.}
\vspace*{0.5ex}
\begin{tabular}{lcccl}
\hline
Decay ~~&~~~~~  $N_{\rm sig}$    ~~~~~&~~~~~   {\large $\epsilon_{\rm av}$}[\%] ~~~~~&~~ $S$[$\sigma$]     ~~~~&~~~~~ ${\cal B}$              \\
%      &      & [\%]       & signif. [$\sigma$] &~~ fraction \\
\hline
$B^0\to D_s^-(\to \phi\pi^-) K^0_S\pi^+$ & $34.6^{+7.1}_{-6.3}$ & $9.09 \pm 0.19$ & $7.4$ &
$0.37 \pm 0.08$~~~~~~~~~~~~~ \\
$B^0\to D_s^-(\to K^{*0} K^-) K^0_S\pi^+$ & $32.9^{+8.9}_{-8.2}$ & $5.99 \pm 0.16$ & $4.5$ &
$0.46 \pm 0.13$ $~~\times 10^{-4}$ \\
$B^0\to D_s^-(\to K^0_S K^-) K^0_S\pi^+$ & $29.2^{+7.4}_{-6.7}$ & $8.68 \pm 0.29$ & $5.7$ &
$0.72 \pm 0.18$~~~~~~~~~~~~~ \\
 &   &   \multicolumn{3}{c}{simultaneous:} \\
 &   &   & 10.1  &   $0.47 \pm 0.06 \pm 0.05$ \\
\hline
$B^+\to D_s^-(\to \phi\pi^-) K^+K^+$ & $15.2^{+5.0}_{-4.3}$ & $11.62 \pm 0.14$ & 5.1 &  $0.87 \pm 0.29$~~~~~~~~~~~~~~\\
$B^+\to D_s^-(\to K^{*0} K^-) K^+K^+$ & $3.8^{+4.7}_{-3.8}$ & $10.22 \pm 0.13$  & 1.0 &  $0.22 \pm 0.31$ $~~\times 10^{-5}$\\
$B^+\to D_s^-(\to K^0_S K^-) K^+K^+$ & $21.5^{+6.5}_{-5.7}$ & $12.11 \pm 0.29$ & 5.2 &  $2.64 \pm 0.78$~~~~~~~~~~~~~~\\
 &   &    \multicolumn{3}{c}{simultaneous:} \\
 &   &   & 6.6 & $0.93 \pm 0.22 \pm 0.10$ \\
\hline
%\cline{1-3}
\end{tabular}
\label{RESULTS}
\end{center}
\end{table*}

An additional background component $j={B^0}^{{\rm bkg}}$  ($j=D_s^{(*)}K\pi$) is introduced for $D_s^-K^0_S \pi^+$ ($D_s^-K^+K^+$), according to the results of dedicated MC studies.
For the $B^0 \to D_s^-K^0_S \pi^+$ decay, we define
\begin{equation}
{\cal P}^{{B^0}^{{\rm bkg}}}_i = {\cal G}(\Delta E^i;~\overline{\Delta E}, \sigma_{\Delta E}) \times {\cal G}(M^i_{\rm bc};~ m_B, \sigma_{M_{\rm bc}}) \times p_2(M^i(D_s);~v_0,v_1,v_2),
\label{EQ5}
\end{equation}
to model the $B^0$ peaking background. For the $B^+ \to D_s^-K^+K^+$ channel, the respective background PDF contribution is defined by
%%%%%%%%%%%
%
\begin{equation}
\begin{split}
{\cal P}^{D_s^{(*)}K\pi}_i &= \Big[f^{D_sK\pi} \cdot {\cal G}_b (\Delta E^i;~\overline{\Delta E}^b, \sigma_{\Delta E}^{b1}, \sigma_{\Delta E}^{b2}) + (1 - f^{D_sK\pi}) \cdot {\cal C}(\Delta E^i;~\overline{\Delta E}^{\cal C}, \sigma^{\cal C}, \alpha^{\cal C}, n^{\cal C})\Big] \\
&\times \Big[f^{D_sK\pi} \cdot {\cal G}(M^i_{\rm bc};~ m_B, \sigma_{M_{\rm bc}}) + (1 - f^{D_sK\pi}) \cdot {\cal G}_b (M^i_{\rm bc};~ m_B^b, \sigma^{b1}_{M_{\rm bc}},\sigma^{b2}_{M_{\rm bc}} )\Big] \\
& \times {\cal G}_2\bigl(M^i(D_s);~m_{D_s}, \sigma^{(1)}_{D_s}, \sigma^{(2)}_{D_s}, f^{\rm sig}_{D_s}\bigr),
\end{split}
\end{equation}
where a bifurcated Gaussian (${\cal G}_b$) and a Crystal Ball function (${\cal C}$)\cite{CRYSTALBALL} are used to parameterize the $B^+ \to D_s^{(*)-} K^+\pi^+$ component. The relevant parameters ($\overline{\Delta E}^b, \sigma_{\Delta E}^{b1}, \sigma_{\Delta E}^{b2}, m_B^b, \sigma^{b1}_{M_{\rm bc}}, \sigma^{b2}_{M_{\rm bc}}$  for $\mathcal{G}_b$  and $\overline{\Delta E}^{\cal C}, \sigma^{\cal C}, \alpha^{\cal C}, n^{\cal C}$ for $\mathcal{C}$) are fixed from a fit to the $B^+ \to D_s^{(*)-} K^+\pi^+$ MC samples; $f^{D_sK\pi}$, the relative contribution of $D_sK\pi$ and $D_s^*K\pi$ events, is evaluated from the $D_s K\pi$ and $D_s^{*} K\pi$ MC samples for each $D_s$ mode. The values of the remaining quantities are treated in a fashion similar to that of the $B^0 \to D_s^-K^0_S \pi^+$ channel. The obtained signal yields ($N_{\rm sig}$) are listed in Table~\ref{RESULTS}.
\begin{table*}[t]
\begin{center}
\caption{Systematic uncertainties (in \%) on the branching fractions for $B^0 \to D_s^-K^0_S \pi^+$ and $B^+ \to D_s^-K^+K^+$ decay modes.} 
\vspace*{0.5ex}
\begin{tabular}{lcc}
\hline
{Source} &  
{$B^0 \to D_s^-K^0_S \pi^+$}~~ & ~~
{$B^+ \to D_s^-K^+K^+$} \\
\cline{2-3} 
\hline
(a)~  Selection procedure            & $\pm$3.6 & $\pm$3.6  \\
(b)~  Background components          & -3.4 & +1.7  \\
(c)~  Signal shape                   & $\pm$3.4  & $\pm 4.6$ \\
(d)~  MC statistics and fit bias     & $\pm 2.8$  & $\pm 2.9$ \\
(e)~  ${\cal B}_{int}$               &  $\pm 5.2$ &  $\pm 5.2$ \\
(f)~  Tracking                       & $\pm$3.6 & $\pm$4.6  \\
(g)~  Hadron identification          & $\pm$3.1 & $\pm$4.9  \\
(h)~  $K^0_S$ reconstruction         & $\pm$5.9 & $\pm$1.0 \\
(i)~~  Uncertainty in N($B\overline{B}$) & $\pm$1.4 & $\pm$1.4 \\
\hline
Total  & $\pm 11.3$ & $\pm 11.0$ \\
\hline
\end{tabular}
\label{SYST}
\end{center}
\end{table*}
Figures~\ref{FIG_DSKSPI} and~\ref{FIG_DSKK} show the distributions of  $\Delta E$, $M_{\rm bc}$ and
$M(D_s)$ for $B^0 \to D_s^-K^0_S \pi^+$ and $B^+ \to D_s^-K^+K^+$, respectively, together with the fits described above. 

We study the invariant mass distribution of the $D_s^-K^0_S$ ($D_s^-K_{\rm low}^+$) subsystem in the $D_s^-K^0_S \pi^+$ ($D_s^-K^+K^+$) final state, where $K_{\rm low}^+$ is the kaon with the lower momentum. 
These distributions exhibit a surplus in the low $D_s K$ mass region with enhancements around $2.7\,\textrm{GeV}/c^2$~(Fig.~\ref{DSKPAIR}).
For each $D_s$ decay mode in both channels, we obtain the respective branching fraction (${\cal B}$) by performing another fit by substituting $N_{\rm sig}$ in Eq.(1) with
\begin{equation}
N_{\rm sig} = {\cal B} \cdot \epsilon(M(D_sK)) \cdot N_{B\bar{B}} \cdot {\cal B}_{\rm int},
\label{BF_form}
\end{equation}
where $N_{B\bar{B}}$ denotes the number of $B$ meson pairs in the data sample and ${\cal B}_{\rm int}$ is the product of branching fractions for the decays of the intermediate resonances in the respective decay chain.
To account for efficiency variations for observed data,
we use an efficiency $\epsilon(M(D_sK))$ that is measured in bins of $M(D_sK)$.
The combined branching fraction is calculated by performing a simultaneous fit to the three $D_s^-$ decay modes using a common ${\cal B}$ value.

The average reconstruction efficiencies ({\large $\epsilon_{\rm av}$}), branching fractions and the signal yields, together with their statistical significances ($S$), are listed in Table~\ref{RESULTS}. The significance is defined as $\sqrt{-2 {\rm ln}({\cal L}_0/{\cal L}_\mathrm{max})}$, where ${\cal L}_\mathrm{max}$ (${\cal L}_0$) denotes the  maximum likelihood with the signal yield at its nominal value (fixed to zero).
The {\large $\epsilon_{\rm av}$} values are calculated from Eq.(\ref{BF_form}) using the obtained $N_{\rm sig}$ and ${\cal B}$ values for each channel, where $\epsilon(M(D_sK))$ is replaced by {\large$\epsilon_{\rm av}$}. 
The systematic uncertainties, described below, are evaluated for the full data sample for all three $D_s$ decay modes.

Systematic uncertainties are listed in Table~\ref{SYST}.
The contribution due to the selection procedure, item (a), is dominated by the $R_2$ requirement. It is estimated in the control channel by comparing the signal ratios for the data and dedicated MC sample. Each ratio is constructed by dividing the nominal signal yield by that without the $R_2$ requirement. 
The uncertainty due to the background components (b) for $B^0 \to D_s^-K^0_S \pi^+$ decay is determined by studying the possible influence of the low-$\Delta E$ region on the signal yield by adding the respective component to the PDF, which includes a peaking background in the $M_{\rm bc}$ and $M_{D_s}$ variables. For $B^+ \to D_s^-K^+K^+$, we compare the nominal branching fraction with the one obtained from the fit with the $B^+ \to D_s^{*-} K^+\pi^+$ component ignored in the PDF. 
To evaluate the contribution related to the signal shape (c), we repeat the fits while varying the fixed shape parameters by $\pm 1\sigma$.
The uncertainty due to limited MC statistics (d) is dominated by the statistical error on the selection efficiency. It is evaluated by varying the $\epsilon(M(D_sK))$ values within their statistical errors in the efficiency distributions over $M(D_sK^0_S)$ and $M(D_sK)$ and comparing the modified branching fractions with the nominal values. This uncertainty also includes a small contribution from the possible fit bias, which is evaluated by comparing the number of MC signal events with the corresponding value obtained from the fit.
Contribution (e) is due to uncertainties in the branching fractions for the decays of intermediate particles, predominantly those of the $D_s$~\cite{PDG}.
The overall
systematic error is obtained by summing all contributions in
quadrature.

Using the branching fraction for the $B^+ \to D_s^-K^+\pi^+$ decay~\cite{MOJE} obtained with a method similar to that of the $B^+ \to D_s^-K^+K^+$ studies, we calculate the ratio
\begin{equation}
{\cal R}_{\cal B} \equiv \frac{{\cal B}(B^+ \to D_s^-K^+K^+)}{{\cal B}(B^+ \to D_s^-K^+\pi^+)} 
= 0.054 \pm 0.013 ({\rm stat}) \pm 0.006 ({\rm syst}),
\end{equation}
where the common systematic uncertainties cancel. The value of the ratio is consistent with the theoretical expectation from the na\"{\i}ve factorization model, 
\begin{equation}
{\cal R}_{\cal B}^{th} = \left(\frac{|V_{us}|}{|V_{ud}|}\right)^2 \cdot \left(\frac{f_K}{f_{\pi}}\right)^2 \cdot \frac{{\cal V}(D_sKK)}{{\cal V}(D_sK\pi)} 
= 0.066 \pm 0.001,
\end{equation}
where $f_h$ is the decay constant for a given hadron $h$~\cite{PDG} and ${\cal V}(D_sKh$) is the phase-space volume for the respective final state.

In summary, we have determined the following branching fractions:
\begin{equation}
 {\cal B}(B^0 \to D_s^-K^0_S \pi^+) 
 = [0.47 \pm 0.06 ({\rm stat}) ~\pm 0.05 ({\rm syst})]\times 10^{-4}~~~~
\end{equation}
and
\begin{equation}
 {\cal B}(B^+ \to D_s^-K^+K^+)
 = [0.93 \pm 0.22 ({\rm stat}) \pm 0.10 ({\rm syst})]\times 10^{-5}.~~
\end{equation}
They are consistent with, and more precise than, the values reported by the BaBar Collaboration~\cite{BABAR_DSKAPI}.
The comparison of the branching fractions for the Cabibbo-suppressed decay $B^+ \to D_s^-K^+K^+$ to the Cabibbo-favored $B^+ \to D_s^-K^+\pi^+$ process yields a result compatible with the na\"{\i}ve factorization hypothesis.
We also find a deviation from the simple phase-space model in the $D_sK$ invariant-mass distributions for both decays. A similar and significant effect has already been observed in other hadronic~\cite{MOJE, BABAR_DSKAPI} and semileptonic~\cite{JACO} decays.
This phenomenon may be related to strong interaction effects in the $\bar{c}s\bar{s}q$ quark system ($q = d, u$) and, in particular, might be explained by the production of charm resonances with masses below the $D_s^{(*)}K$ threshold~\cite{THEOR}. A more detailed analysis of the enhancement (\textit{e.g.,} a study of the angular distribution) requires larger data samples that will be accessible to the LHCb~\cite{LHCB} and Belle II~\cite{BELLE2} experiments. 

%%%%%%%%%%%%%%%%%%%%%%%%%%%%%%%%%%%%%%%%%%%%%%%%%%%%%%%%%%%%%%%%%%%%%%%%%%%%%%
We thank the KEKB group for the excellent operation of the
accelerator; the KEK cryogenics group for the efficient
operation of the solenoid; and the KEK computer group,
the National Institute of Informatics, and the 
PNNL/EMSL computing group for valuable computing
and SINET4 network support.  We acknowledge support from
the Ministry of Education, Culture, Sports, Science, and
Technology (MEXT) of Japan, the Japan Society for the 
Promotion of Science (JSPS), and the Tau-Lepton Physics 
Research Center of Nagoya University; 
the Australian Research Council and the Australian 
Department of Industry, Innovation, Science and Research;
Austrian Science Fund under Grant No.~P 22742-N16 and P 26794-N20;
the National Natural Science Foundation of China under Contracts 
No.~10575109, No.~10775142, No.~10825524, No.~10875115, No.~10935008 
and No.~11175187; 
the Ministry of Education, Youth and Sports of the Czech
Republic under Contract No.~LG14034;
the Carl Zeiss Foundation, the Deutsche Forschungsgemeinschaft
and the VolkswagenStiftung;
the Department of Science and Technology of India; 
the Istituto Nazionale di Fisica Nucleare of Italy; 
National Research Foundation of Korea Grants
No.~2011-0029457, No.~2012-0008143, No.~2012R1A1A2008330, 
No.~2013R1A1A3007772, No.~2014R1A2A2A01005286, No.~2014R1A2A2A01002734, 
No.~2014R1A1A2006456;
the BRL program under NRF Grant No.~KRF-2011-0020333, No.~KRF-2011-0021196,
Center for Korean J-PARC Users, No.~NRF-2013K1A3A7A06056592; the BK21
Plus program and the GSDC of the Korea Institute of Science and Technology Information;
the Polish Ministry of Science and Higher Education and 
the National Science Center;
the Ministry of Education and Science of the Russian
Federation and the Russian Federal Agency for Atomic Energy;
the Slovenian Research Agency;
the Basque Foundation for Science (IKERBASQUE) and the UPV/EHU under 
program UFI 11/55;
the Swiss National Science Foundation; the National Science Council
and the Ministry of Education of Taiwan; and the U.S.\
Department of Energy and the National Science Foundation.
This work is supported by a Grant-in-Aid from MEXT for 
Science Research in a Priority Area (``New Development of 
Flavor Physics'') and from JSPS for Creative Scientific 
Research (``Evolution of Tau-lepton Physics'').


\begin{thebibliography}{99}
\bibitem{FOOT} 
  Throughout the paper, the inclusion of the charge-conjugate decay mode is implied unless otherwise stated.
\bibitem{MOJE} J. Wiechczynski \textit{et al.} (Belle Collaboration), Phys. Rev. D \textbf{80}, 052005 (2009).
\bibitem{BABAR_DSKAPI} 
  B.~Aubert \textit{et al.} (BaBar Collaboration), Phys. Rev. Lett.  {\bf 100}, 171803 (2008).
\bibitem{FACTOR} M.~J. Dugan and B. Grinstein, Phys. Lett. B \textbf{255}, 583 (1991).
\bibitem{CKM_PAPER1} 
N. Cabibbo, Phys. Rev. Lett. {\bf 10}, 531 (1963).
\bibitem{CKM_PAPER2} 
M. Kobayashi and T. Maskawa, Prog. Theor. Phys. {\bf 49}, 652 (1973).
%%%%%%%%%%%%
\bibitem{JACO} 
J. Stypula \textit{et al.} (Belle Collaboration), Phys. Rev. D \textbf{86}, 072007(R) (2012).


%%%%%%%%%%%%
\bibitem{KEKB}
S.~Kurokawa and E.~Kikutani, Nucl. Instrum. Methods Phys. Res. Sect., 
 A {\bf 499}, 1 (2003), and other papers included in this Volume;
 T.~Abe \textit{et al.}, Prog. Theor. Exp. Phys. (2013) 03A001 and following
 articles up to 03A011.

%%%%%%%%%%%%
\bibitem{BELLE}  
A.~Abashian {\it et al.} (Belle Collaboration), Nucl. Instrum. Methods 
 Phys. Res., Sect. A {\bf 479}, 117 (2002); also see the detector section in
 J.~Brodzicka \textit{et al.}, Prog. Theor. Exp. Phys. (2012) 04D001.

%%%%%%%%%%%%
\bibitem{SVD} 
  %Z.~Natkaniec \textit{et al.} (Belle SVD2 Group), Nucl. Instr. and Meth. A \textbf{560}, 1 (2006).
  Z.~Natkaniec {\it et al.} (Belle SVD2 Group), Nucl. Instrum. Methods Phys. Res., Sect. A \textbf{560}, 1 (2006).
%%%%%%%%%%%%
\bibitem{PDG}  
J. Beringer et al. (Particle Data Group), Phys. Rev. D {\bf 86}, 010001 (2012).


%%%%%%%%%%%%
\bibitem{GEANT}  
  R.~Brun \textit{et al.}, GEANT 3.21, CERN REPORT DD/EE/84-1, 1984.

%%%%%%%%%%%%
\bibitem{FOX} 
  G.~C.~Fox and S.~Wolfram, Phys. Rev. Lett. \textbf{41}, 1581 (1978).

%%%%%%%%%%%%
\bibitem{ARGUS} 
  H.~Albrecht \textit{et al.} (ARGUS Collaboration), Phys. Lett. B \textbf{241}, 278 (1990).


\bibitem{CRYSTALBALL} T.~Skwarnicki, Ph.D. Thesis, DESY F31-86-02(1986), Appendix E.
%%%%%%%%%%%%

\bibitem{THEOR} 
  O.~Antipin and G.~Valencia, Phys. Lett. B \textbf{647}, 164 (2007).

\bibitem{LHCB} A.~A. Alves Jr. {\it et al.}  (LHCb Collaboration), JINST \textbf{3} S08005 (2008).

\bibitem{BELLE2} T. Abe \textit{et al.}, arXiv:1011.0352v1 [physics.ins-det] (2010).

\end{thebibliography}
\end{document}